\documentclass[%
 reprint,
 amsmath,amssymb,
 aps,
]{revtex4-2}

\usepackage{graphicx}
\usepackage{dcolumn}
\usepackage{bm}
\usepackage{xcolor,soul,framed} 
\usepackage{subfigure}
\usepackage{fancyhdr}
\usepackage{booktabs}
\usepackage{tabularx}
\usepackage{comment}

\newcommand{\red}[1]{\textcolor{red}{#1}}
\newcommand{\blue}[1]{\textcolor{blue}{#1}}

\newcommand{\Fig}[1]{Figure\,{\ref{#1}}}
\newcommand{\Figs}[1]{Figures\,{\ref{#1}}}
\newcommand{\E}{\mathrm{e}}
\newcommand{\jj}{\mathrm{j}}

\begin{document}


\title{{Analytical Equivalent Circuits for Three-dimensional Metamaterials and Metagratings}}

\

\author{Antonio Alex-Amor$^{*}$}
\affiliation{
Department of Information Technology, 
Universidad San Pablo-CEU, CEU Universities,  Campus Montepríncipe, 28668 Boadilla del Monte (Madrid), Spain
}%
\email{Corresponding author: antonio.alexamor@ceu.es}
\author{Salvador Moreno-Rodríguez, Pablo Padilla, Juan F. Valenzuela-Valdés, Carlos Molero}
\affiliation{%
Department of Signal Theory, Telematics and Communications, Research Centre for Information and \\Communication Technologies (CITIC-UGR), University of Granada, Granada, Spain
}%

\begin{abstract}
In recent times, three-dimensional (3D) metamaterials have undergone a revolution driven mainly by the popularization of 3D-printing techniques, which has enabled the implementation of modern microwave and photonic devices with advanced functionalities. However, the analysis of 3D metamaterials is complex and computationally costly in comparison to their 1D and 2D counterparts due to the intricate geometries involved.  In this paper, we present a fully-analytical framework based on Floquet-Bloch modal expansions of the electromagnetic fields and integral-equation methods for the analysis of 3D metamaterials and metagratings. Concretely, we focus on 3D configurations formed by periodic arrangements of rectangular waveguides with longitudinal slot insertions. The analytical framework is computationally efficient compared to full-wave solutions and also works under oblique incidence conditions. Furthermore, it comes associated with an equivalent circuit that allows to gain physical insight into the scattering and diffraction phenomena. The analytical equivalent circuit is tested against full-wave simulations in commercial software CST. Simulation results show that the proposed 3D structures provide independent polarization control of the two orthogonal polarization states. This key property is of potential interest for the production of full-metal polarizers, such as the one illustrated here.
\end{abstract}

\maketitle


\section{\label{sec:introduction}Introduction}

Modern analog microwave and photonic devices, such as frequency selective surfaces (FSS),  waveguide devices, filters, absorbers, antennas or polarizers, are based on the use of metamaterials  \cite{Marcuvitz86, Vengsarkar1996, Kildal2009, Optical_polarizer, SailingHe2012, Minatti2015, Palomares2020, Quevedo2016}. Metamaterials are human-made artificially-engineered devices that allow arbitrary control and manipulation of the propagation of electromagnetic waves \cite{EnghetaBook2006, CapolinoBook2006}. Traditionally, metamaterials have rested on periodic or quasi-periodic arrangements of subwavelength insertions, that is, structures whose constituent elements repeat periodically in space \cite{Pendry2004}, time \cite{Galiffi2022} or space-time \cite{Caloz2020_1, Taravati2019}. Historically, scientific and engineering communities have paid special attention to one-dimensional (1D) and two-dimensional (2D) metamaterial configurations due to the simplicity related to their analysis, design and manufacturing \cite{QuevedoRoadmap2019}. Nonetheless, 1D and 2D metamaterials present fundamental limitations inherent to their geometry, such as independent orthogonal polarization control \cite{Alex3D_2022}. These limitations are mainly due to the fact that 1D and 2D configurations do not exploit the degrees of freedom associated to the longitudinal direction, the spatial direction in which the wave actually propagates. In that sense, 3D metamaterials are called to overcome the limitations of 1D and 2D configurations \cite{3Dreview2019}, leading to a new era in the metamaterial field from which wireless communication systems can benefit. 

The popularization of 3D metamaterials is recent. At the cost of increasing the cell thickness, a new degree of freedom is introduced, which can be used to improve the performance of the device. This is associated to the exploitation of the longitudinal direction ($z$ direction) in design. The different homogeneous longitudinal  sections, i.e. waveguide regions, can be modified with the insertion of longitudinal slots. The inclusion of longitudinal slots allow us to tune the electromagnetic response of the 3D device in an effective manner, with potential application in reflectarray and transmitarray technology and other microwave and photonic devices \cite{Alex3D_2022, 3Dreview2019}.  For instance, this inclusion often increases angular stability, allows dual-band frequency responses,  enhances the operational bandwidth and provides structural robustness. This should be considered a valuable asset not present in flat 2D metasurfaces or in stacks of 2D devices (2.5D metastructures).


For the aforementioned reasons, 3D metamaterials are starting to be applied in microwave and photonics engineering for the realization of advanced polarizers \cite{Molero3D_2020}, absorbers \cite{Tsilipakos2020}, beamforming systems \cite{Molero2021}, wide-angle impedance matching layers \cite{Diego2022} and to control orthogonal linear polarizations in reflectarray/transmitarray cells \cite{Palomares2023}. This has been made possible thanks to the evolution of 3D-printing techniques and the impressive increase in computational resources in the last years \cite{Alex3D_2022, 3Dreview2019, Hoon2020, SanchezOlivares2022, Garcia-Vigueras22}.  Nonetheless, 3D metamaterials are bulkier than flat devices  and, usually, more difficult to be analyzed due to their complex geometry. Robust and generalist full-wave tools, such as the finite elements method (FEM) or finite-difference time-domain (FDTD) techniques \cite{Itoh89}, can be employed for the analysis of 3D metamaterials. Normally, the use of the previously mentioned methods comes at the price of great computational resources and a lack of physical insight into the electromagnetic behavior of the structure. Some more efficient and physically-insightful alternatives to full-wave methods were discussed in \cite{Alex3D_2022}, among which homogenization theory \cite{Blanchard1994, Silveirinha2005}, modal analysis \cite{Epstein2020, Kari2020}, ray optics \cite{Deschamps1972, Liao2023}, transfer-matrix analysis \cite{Li2003, Giusti2022}, circuit models
\cite{GrbicCircuit2005, Zedler2007} or some specific combination of these can be found. 


Among the methods mentioned above, circuit models are of particular interest. Complex physical phenomena can be described in a straightforward manner with the use of equivalent circuits ruled by basic circuit theory and its main components: voltage/current sources, impedances, admittances and transmission lines \cite{Mesa2018, Costa2012}. Furthermore, circuit models are remarkably more computationally efficient than other numerical techniques. Without loss of generality, we can classify equivalent circuits into two main types: (i) heuristic and (ii) analytical. Heuristic approaches need of the support of an external method or simulator, such as CST Studio or Ansys HFSS, to calculate the value of their circuit components. There exist many examples in the literature where heuristic equivalent circuits are utilized to model and characterize complex electromagnetic phenomena in metamaterials and FSS structures \cite{Kafesaki2007, Carbonell2010, PerezPalomino2018, Borgese2020, Esteban2021}.  On the other hand, analytical circuit approaches do not need of external support, i.e., they are fully operational by themselves \cite{Dubrovka2006, Torres2016, Khavasi2014, Mesa2016, Khavasi2015}. This fact constitutes a major difference between heuristic and analytical equivalent circuits. Nevertheless, analytical circuit models are often restricted to canonical geometries since complex geometries may not be easily modeled with analytical mathematical expressions.

In this paper, we propose a rigorous and systematic analytical framework based on integral-equation techniques and Floquet-Bloch series expansions of the electromagnetic fields to analyze 3D metamaterials and metagratings. Related formulations have been successfully applied for the analysis of 1D \cite{Berral2012_1, Berral2012, Molero2014, Molero2017_asymmetrical} and 2D \cite{Dubrovka2006, Berral2015, Meander2017, Alex2021, FloquetCircuit_2D2} metamaterial structures in the past and, more recently, to time-varying systems \cite{Alex2023, Salva2023, AlexLorentz2023}. Similarly to previous approaches, the analytical formulation is connected to an \emph{analytical equivalent circuit} that models the 3D structure. 

In all cases, we are dealing with thick (nonflat) metamaterials with a 3D profile, formed by 2D-periodic arrangements of slotted waveguides. The insertion of longitudinal slots modify the electromagnetic response of the 3D metastructure, allowing us to control and manipulate the transmission and reflection of electromagnetic waves. The proposed 3D metastructures are fundamentally based on metallic waveguide geometries. Thus, their interior is essentially hollow, filled with air. This causes the weight of the structure to be reduced, since the volume of metal or metallized material is small compared to the total volume of the metadevice.

The equivalent circuit is constituted by lumped elements that describe higher-order wave coupling between the different waveguide sections, and transmission lines that characterize the wave propagation in the different regions. To the best of our knowledge, this is one the first times that a rigorous fully-analytical equivalent circuit is proposed to model a 3D metamaterial.  Some of the previous approaches found in the literature show pure heuristic \cite{GrbicCircuit2005, Diego2022, Balmaseda2023} or quasi-heuristic (mixture of heuristic and analytical) \cite{Carlos_MTT3D} rationales, but none is purely analytical. Fully-analytical schemes are preferred over heuristic or quasi-heuristic ones, as they are independent of external full-wave simulations. Thus, the present analytical equivalent circuit can be used as an efficient surrogate model and be combined with artificial intelligence or conventional optimization techniques for the design of 3D devices.

All the considered structures are of fully-metallic nature. Nonetheless, the inclusion of dielectrics in the hollow waveguide sections can be easily treated from an analytical perspective with the circuit approach, if necessary. There are plenty of commercial applications where fully-metallic structures are preferred over dielectric ones. For instance,  the use of dielectrics is not recommended in space applications, as the thermal expansion coefficients of dielectrics differ from those of metals, resulting in an uneven, non-uniform volume expansion, which can lead to structure failure. In general, fully-metallic structures fit very well in scenarios where systems are subjected to large thermal variations, both in space and on Earth. Additionally, the use of fully-metallic configurations, such as the ones presented in this work, is beneficial for operation at high frequencies. This is motivated by two main reasons. Firstly, dielectrics increase structure losses significantly as frequency increases. Fully-metallic structures are much more robust to losses, as ohmic losses are much easier to control provided that proper fabrication processes are applied. Secondly, fully-metallic designs can be easily scaled in frequency compared to  mixed metallo-dielectric or fully-dielectric designs.  

The paper is organized as follows. Section II presents the derivation of the analytical equivalent circuit that models 3D metamaterials. Then, some numerical computations are performed in order to check the correct operation of the circuit approach, including reflective and transmitting structures. Section III shows a relation between the original 3D metagrating and related configurations. It is shown that, under certain circumstances, related advanced configurations can be also analyzed with the present approach. Section IV details the utilization of the present circuit model for the efficient design of polarizer devices. Finally, Section V presents some general conclusions extracted from the work.

\section{\label{sec:Theory} Theoretical Framework \& Results}

\begin{figure*}[!t]
	\centering
	\subfigure[]{\includegraphics[width= 0.65\columnwidth]{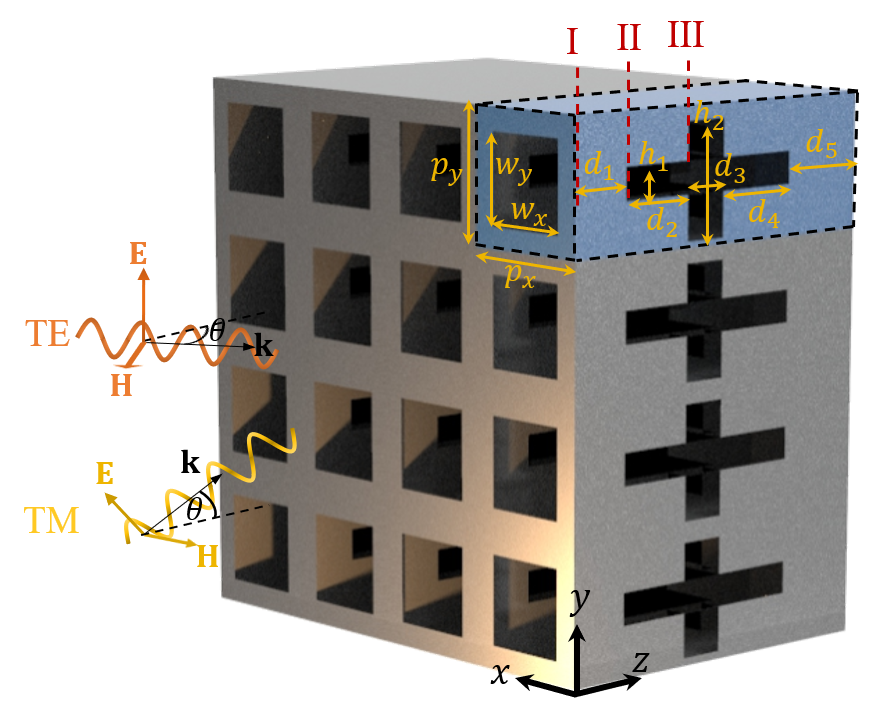}}
    \hspace*{2cm}
     \subfigure[]{\includegraphics[width= 0.6\columnwidth]{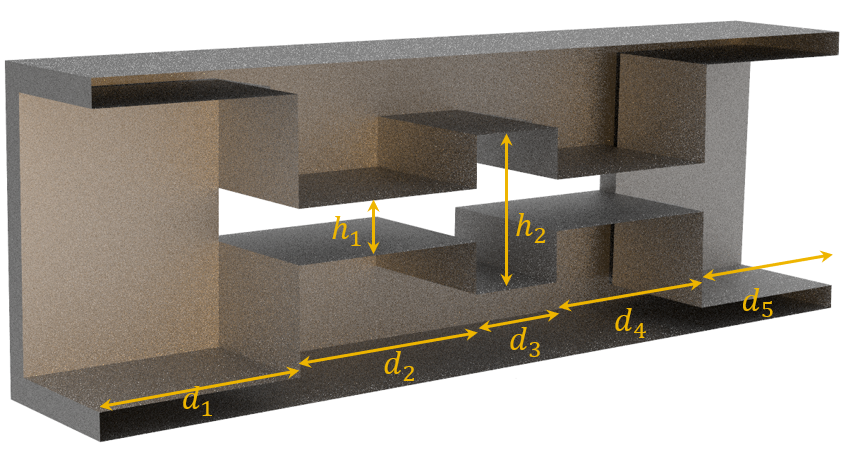}} \\ 
    \subfigure[]{\vspace{2.5cm} \includegraphics[width= 1.8\columnwidth]{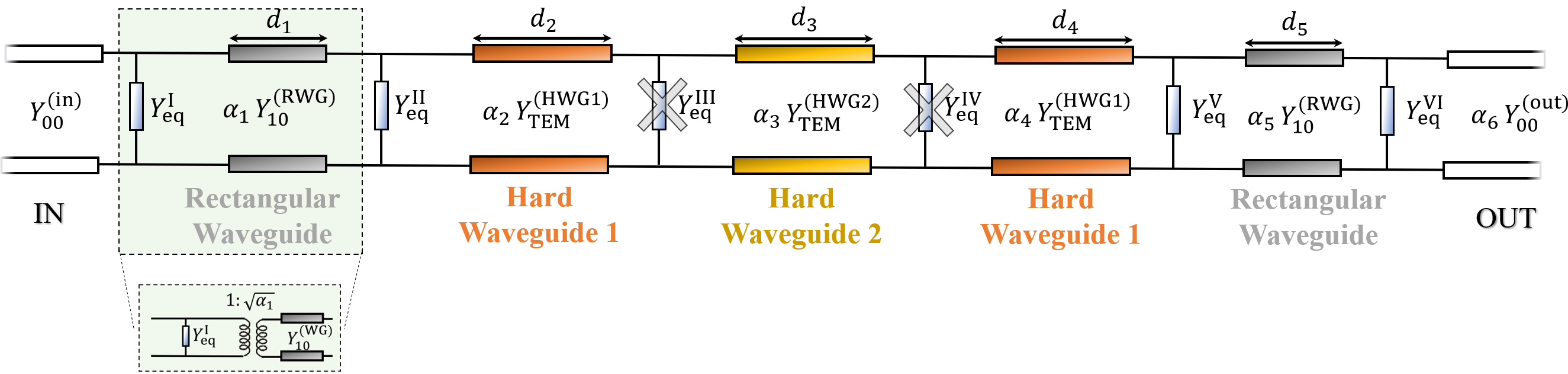}}
	\caption{(a) 3D metamaterial formed by periodic repetitions of rectangular waveguides with longitudinal cross-shaped  slot insertions.  The unit cell of the 3D structure is highlighted in blue. (b) Cross-sectional view of a unit cell. (c) Analytical Floquet equivalent circuit.}
	\label{fig1}
\end{figure*}

The original metamaterial under consideration is illustrated in \Fig{fig1}(a). It is a thick 3D metastructure formed by a 2D-periodic arrangement (periodicities $p_{\text{x}}$ and $p_{\text{y}}$ along the $x$ and $y$ directions, respectively) of metallic waveguides with slot insertions placed along the longitudinal direction $z$. The unit cell of the 3D metamaterial is highlighted in blue and the main geometrical parameters are also remarked. The corresponding unit cell is bounded by periodic boundary conditions (PBCs) in the $x$ and $y$ directions. 

The 3D metagrating (with 2D periodicity) shown in \Fig{fig1} share some similarities with multilayer devices formed by stacks of 2D metasurfaces (commonly named as 2.5D structures), especially from an analytical perspective. In 2.5D structures, each (meta)layer that form the stack is homogenized and modeled as a shunt impedance/admittance, while the connecting  homogeneous media between layers (typically air or generic dielectrics) is modeled as transmission lines \cite{Alex2021, FloquetCircuit_2D2}.  A similar approach is followed here for the analysis of the proposed 3D metadevice. The transitions between different waveguide sections are modeled with shunt admittances, while the connecting media (homogeneous waveguides) with transmission lines. Although the applied rationale could seem to be rather similar in both 2.5D and 3D cases, the homogeneous waveguide sections discussed here are significantly more complex to be analytically described, as well as the transitions between them. More importantly, the main difference between a 2.5D device (multilayer stack) and the proposed 3D device lies in the fact that, in a multilayer stack, the longitudinal direction cannot be exploited from a design perspective. This is not the case in the 3D device shown here, where the homogeneous waveguide sections can be modified with the insertion of longitudinal slots, which allows us to control and manipulate in a more efficient manner the transmission and reflection of electromagnetic waves.

Thus, the longitudinal structure of the cell can be split in several homogeneous regions. On the one hand, regions where there are no perforations, emulating conventional metallic rectangular waveguides (RWG) with dimensions $w_{\text{x}} \times w_{\text{y}}$. On the other hand, regions where perforations exist, emulating homogeneous \emph{hard waveguides} (HWGs) with dimensions $p_{\text{x}} \times h_{1,2}$, and where the TEM mode can propagate. 
As the cross-sectional view in \Fig{fig1}(b) shows, HWG regions are stretched along the $y$ axis. In general, HWGs can be regarded as parallel-plate waveguides whose lateral walls are periodic boundary conditions. When normal incidence is considered, the symmetry of the cell with respect the principal planes reduces the periodic boundary conditions on the walls to  perfect magnetic/electric conductors (PMC or PEC) conditions. For  an incident electric-field vector polarized along $\hat{\mathbf{y}}$ (according to the frame of coordinates in \Fig{fig1}), PBCs in the YZ plane become PMCs. This situation is the most interesting since slot resonators may exhibit their resonant conditions. We will henceforth focus on this scenario.

The equivalent circuit that describes the physical phenomenology associated to the 3D metagrating is illustrated in \Fig{fig1}(c). Each discontinuity is modeled as a shunt equivalent admittance that takes into account all relevant information about higher-order coupling between evanescent modes/harmonics. For the present 3D structure, we have three main discontinuities, labeled as I, II, III. Discontinuity I models the transition between the input media (typically considered here to be air) and the RWG. Discontinuity II models the transition between the RWG and the lowest hard waveguide (WG--HWG transition). Discontinuity III models the transition between the lowest and the highest hard  waveguides (HWG--HWG transition). Discontinuities IV, V and VI are of the same type than discontinuities III, II and I, respectively. This will be discussed later in more detail. Finally,  each waveguide section (rectangular and hard) is circuitally modeled as a transmission line of length $d_i$, with specific values of the characteristic admittance and propagation constant (they depend on the waveguide nature). It is always assumed that the propagation is carried out by the fundamental mode/harmonic of each of the regions. Input (in) and output (out) media are semi-infinite spaces characterized by semi-infinite transmission lines with characteristic admittance $Y^{(\mathrm{in})}_{00}$ and $Y^{(\mathrm{out})}_{00}$, respectively. 

Circuit parameters will be dependent on the geometry of the 3D structure. As will be explained below, the admittances associated with the rest of transmission lines (and other shunt admittances taking part in the equivalent circuit in \Fig{fig1}) are multiplied by a factor $\alpha_i$, denoting the degree of coupling at the discontinuity planes. This factor is circuitally interpreted as transformer with turn ratio $1:\sqrt{\alpha_i}$, as explicitly shown in \cite{Molero2017_asymmetrical, Molero2016}, among others. The current circuit version sketched in \Fig{fig1}(c) is equally valid and has now been selected in order to reduce the matrix-formalism complexity when cascading different transmission line sections (see Appendix A).

As was reported in \cite{Carlos_MTT3D, Molero3D_2020}, slots in the YZ plane are easily excitable by $\hat{\mathbf{y}}$-polarized waves, and exhibit no effects for the opposite polarization. An identical rationale can be employed for resonators on the XZ plane and $\hat{x}$-polarized incident waves. We henceforth focus on cells with resonators on the YZ-plane and fed by $\hat{\mathbf{y}}$-polarized plane waves. The conclusions extracted are valid for the resonators on the XZ-plane fed by $\hat{\mathbf{x}}$-polarized incident electric fields.

\subsection{Input Media -- Waveguide Discontinuity (\textrm{I})} \label{air_SWG}

For the derivation of the circuit parameters, we will consider the incident of a plane wave having the transverse electric-field vector oriented towards $\hat{\mathbf{y}}$. This scenario can be covered by TM ($E_y, E_z, H_x$) and TE ($E_y, H_x, H_z$) polarized plane waves impinging obliquely (angle $\theta$), as illustrated in \Fig{fig1}(a). Time-harmonic variation ($\E^{\jj\omega t}$) is present in all the considered electromagnetic fields and thus suppressed from now onwards. 

The characterization of the discontinuity departs from the knowledge, a priori, of the electromagnetic field expansion at both sides of the discontinuity. In the input region, referenced by the superscript $(\text{in})$, the transverse electric field can be expressed by a Floquet series of harmonics. Each of these harmonics is excited after the interaction of the incident wave and the discontinuity.  
 
Assuming TM incidence, the field expansion in the air-region can be written at the discontinuity plane $(z = 0)$ as follows \cite{Berral2015, Alex2021}:
\begin{multline}\label{E1}
\mathbf{E}^{(\text{in})} (x, y) = \frac{1}{\sqrt{p_{\text{x}} p_{\text{y}}}}(1 + E_{00}^{\text{TM,(in)}} ) \text{e}^{-\text{j} k_{\text{t}} y }\hat{\mathbf{y}} \\ + \frac{1}{\sqrt{p_{\text{x}} p_{\text{y}}}} \displaystyle\sum_{\forall n,m \neq 0,0 
} \hspace*{-0.3cm} E_{nm}^{\text{TE}, (\text{in})} \frac{k_{m} \hat{\mathbf{x}} - k_{n} \hat{\mathbf{y}}}{k_{nm}} \text{e}^{- \text{j} (k_n x + k_m y)}  
\\ 
+ \frac{1}{\sqrt{p_{\text{x}} p_{\text{y}}}} \displaystyle\sum_{\forall n,m \neq 0,0 
} \hspace*{-0.3cm} E_{nm}^{\text{TM}, (\text{in})} \frac{k_{n} \hat{\mathbf{x}} + k_{m} \hat{\mathbf{y}}}{k_{nm}} \text{e}^{- \text{j} (k_n x + k_m y)}
\end{multline}
with the unity being the normalized amplitude of the incident wave and $E_{00}^{\text{TM,(in)}}$ the unknown reflection coefficient. The coefficients $E_{nm}^{\text{TE/TM}, (\text{in})}$ are the unknown amplitudes associated with TE/TM $nm$-harmonics. The transverse wavevector of a $nm$-th harmonic is referred as $k_{nm}$, defined as 
\begin{equation}
k_{nm} = \sqrt{k_{n}^{2} + k_{m}^2}
\end{equation}
with
\begin{align} \label{knTM}
k_{n} &= \frac{2 \pi n}{p_{\text{x}}} \\
\label{kmTM} k_{m} &= \frac{2 \pi m}{p_{\text{y}}} + k_{\text{t}} \,.
\end{align}
The longitudinal component of the wavevector of a $nm$-harmonic is denoted by $\beta_{nm}$  
\begin{equation}\label{beta}
\beta_{nm} = \sqrt{ k_{0}^2 - k_{nm}^2}\,.
\end{equation}
with $k_0 = \omega / c$. 

Similarly, the magnetic field expansion admits to be expressed as follows \cite{Berral2015, Alex2021}:
\begin{multline}\label{H1}
\mathbf{H}^{(\text{in})} (x, y) = -\frac{Y_{00}^{\text{TM,(in)}}}{\sqrt{p_{\text{x}} p_{\text{y}}}}(1 - E_{00}^{\text{TM,(in)}}) \text{e}^{-\text{j} k_{\text{t}} y }\hat{\mathbf{y}} \\ - \frac{1}{\sqrt{p_{\text{x}} p_{\text{y}}}} \displaystyle\sum_{\forall n,m \neq 0,0 
} \hspace*{-0.3cm} Y_{nm}^{\text{TE}, (\text{in})} E_{nm}^{\text{TE}, (\text{in})} \frac{k_{m} \hat{\mathbf{y}} + k_{n} \hat{\mathbf{x}}}{k_{nm}} \text{e}^{- \text{j} (k_n x + k_m y)}
\\ 
- \frac{1}{\sqrt{p_{\text{x}} p_{\text{y}}}} \displaystyle\sum_{\forall n,m \neq 0,0 
} \hspace*{-0.3cm} Y_{nm}^{\text{TM}, (\text{in})} E_{nm}^{\text{TM}, (\text{in})} \frac{k_{n} \hat{\mathbf{y}} - k_{m} \hat{\mathbf{x}}}{k_{nm}} \text{e}^{- \text{j} (k_n x + k_m y)} 
\end{multline}
with
\begin{align}
Y_{nm}^{\text{TE},(\text{in})} &= \frac{\beta_{nm}^{(\text{in})}}{\eta_{0} k_{0}} \\
Y_{nm}^{\text{TM}, (\text{in})} &= \frac{k_{0}}{\eta_{0} \beta_{nm}^{(\text{in})}}
\end{align}
being the $nm$th-order TE and TM admittances, respectively, and $\eta_0$ is the free-space impedance..

In the RWG region, the field expansion is written in terms of the modal solutions of the RWG. However, at frequency ranges far below the excitation of higher-order modes, the electromagnetic field description inside the RWG admits to be represented in terms of a single mode, say the fundamental $\text{TE}_{10}$ mode:
\begin{align}\label{E2}
\mathbf{E}^{\text{(RWG)}}(x, y) &= E_{10}^{\text{TE}, \text{(RWG)}} \frac{2}{\sqrt{2}}\frac{1}{\sqrt{w_{\text{x}} w_{\text{y}}}}  \cos(k_{10}^{\text{(RWG)}} x)  \hat{\mathbf{y}} \\
\label{H2}
\mathbf{H}^{\text{(RWG)}}(x, y) &= Y_{10}^{\text{TE}, \text{(RWG)}} [\mathbf{E}^{\text{(RWG)}}(x, y) \cdot \hat{\mathbf{y}}] \hat{\mathbf{x}}
\end{align}
with $E_{10}^{\text{TE}, \text{(RWG)}}$ the unknown amplitude of the $\text{TE}_{10}$ mode, $k_{10}^{\text{(RWG)}}~=~\pi/ w_{\text{x}}$ and $Y_{10}^{\text{TE, (RWG)}} = \dfrac{\beta_{10}^{\text{(RWG)}}}{\eta_{0} k_{0}}$. The propagation constant $\beta_{10}^{\text{(RWG)}}$ is similarly obtained as in \eqref{beta}, substituting $k_{nm}$ by $k_{10}^{\text{(RWG)}}$.

The analytical derivation of the equivalent circuit demands an \emph{
a priori} estimation of the field profile at the discontinuity. Given the geometry taken into account, the field profile (spatial distribution) is expected to be similar to the fundamental or  $\text{TE}_{10}$ mode at lower frequencies. Thus, the field at the discontinuity admits to be described as 
\begin{equation}\label{profile_TE}
\mathbf{E}_{\text{dis}} = A \cos \left( \frac{\pi}{w_{\text{x}}} x \right) \hat{\mathbf{y}}\,,
\end{equation}
where $A$ is frequency-dependent constant to be determined. This estimation is good enough for frequencies below the onset of higher-order modes inside the RWG. For normal incidence, the second mode excitable is the $\text{TE}_{30}$, whose cutoff frequency $f_{30} = 3f_{10}$,  with $f_{10} = {c\, k_{10}^{(\text{RWG})}} / 2 \pi$ being the cutoff frequency of the $\text{TE}_{10}$ mode.  

The analytical expressions for the unknown amplitudes of each of the modes/harmonics in both regions are obtained after imposing the following conditions at the discontinuity:
\begin{align}\label{EQ1}
\mathbf{E}^{(\text{in})}(x, y) &=  \mathbf{E}_{\text{dis}} \\
\label{EQ2} \mathbf{E}^{\text{(RWG)}}(x, y) &= \mathbf{E}_{\text{dis}} \\
\label{EQ3} \mathbf{E}_{\text{dis}} \times [\mathbf{H}^{\text{(in)}}(x, y)]^{*}  &= \mathbf{E}_{\text{dis}} \times [\mathbf{H}^{\text{(RWG)}}(x, y)]^{*} 
\end{align}
where \eqref{EQ1} and \eqref{EQ2} refer to the continuity of the electric field and \eqref{EQ3} describes the continuity of the Poyinting vector.

By developing the above equations after introducing the harmonic/modal expansion inside, and after several mathematical calculations, we achieve the following expression for the reflection coefficient $E_{00}^{\text{TM, (in)}}$:
\begin{equation}\label{reflection}
E_{00}^{\text{TM,(in)}} = \frac{Y_{00}^{\text{(in)}} - \alpha_{10}^{\text{TE, (RWG)}} Y_{10}^{\text{TE}, \text{(RWG)}} - Y_{\text{eq}}^{\text{I}}}{Y_{00}^{\text{(in)}} + \alpha_{10}^{\text{TE, (RWG)}} Y_{10}^{\text{TE}, \text{(RWG)}} + Y_{\text{eq}}^{\text{I}}}
\end{equation}
with 
\begin{equation}
Y_{\text{eq}}^{\text{I}} = \displaystyle\sum_{\forall n,m \ne 0,0}  \bigg[\alpha_{nm}^{\text{TE, (\text{in})}} Y_{nm}^{\text{TE}, (\text{in})} + \alpha_{nm}^{\text{TM, (in)}} Y_{nm}^{\text{TM}, (\text{in})} \bigg]\,.
\end{equation}
The factors $\alpha_{nm}^{\text{TE/TM, (in)}}$ and $\alpha_{10}^{\text{TE, RWG}}$ represent the coupling among all the harmonics/modes and $\mathbf{E}_{\text{dis}}$, whose expressions are given by:
\begin{align} \label{NTE1_R}
\alpha_{nm}^{\text{TE}, (\text{in})} &= \bigg[\pi ^2 \frac{k_{n}}{k_{nm}}  \frac{\cos(k_{n} w_{\text{x}}/2)}{(k_{n} w_{\text{x}})^2 - \pi^2} \frac{\sin(k_{m} w_{\text{y}}/2)}{ \sin(k_{\text{t}} w_{\text{y}}/2)} \frac{k_{\text{t}}}{k_{m}}\bigg]^2 \\
\label{NTM1_R}
\alpha_{nm}^{\text{TM}, (\text{in})} &=  \frac{k_{m}^2}{k_{n}^2} \alpha_{nm}^{\text{TE}, (\text{in})} \\
\label{NTE2_R}
\alpha_{10}^{\text{TE, (RWG)}} &= \frac{\pi^2}{8}\frac{p_{\text{x}} p_{\text{y}}}{w_{\text{x}} w_{\text{y}}}\bigg[\frac{k_{\text{t}} w_{\text{y}}/2}{\sin(k_{\text{t}} w_{\text{y}}/2)}\bigg]^{2}\,.
\end{align}

The above expressions and the reflection-coefficient equation in \eqref{reflection} leads to the identification of a 
 transmission-line model where $Y_{00}^{\text{in}}$ and $\alpha_{10}^{\text{RWG}} Y_{10}^{\text{TE}, \text{RWG}}$ are the characteristic admittances of the input and output transmission lines in discontinuity I respectively. They are formally the transmission lines in white and grey in the equivalent circuit in \Fig{fig1}. Henceforth, the parameter multiplying the admittance of the $\text{TE}_{10}$ mode is redefined as follows:
 \begin{equation}\label{alpha1_TM}
\alpha_{1} = \alpha_{10}^{\text{TE, (RWG)}}
 \end{equation}
The admittance $Y_{\text{eq}}^{\text{I}}$ is the shunt admittance connecting both transmission lines.

A similar derivation can be done for TE incidence. In this case the transverse electric-field vector points towards $\hat{\mathbf{y}}$ (this plane wave is completed by magnetic field components pointing towards $\hat{\mathbf{x}}$ and $\hat{\mathbf{z}}$). The rationale employed is identical, but some of the expressions must now be redefined. Now, \eqref{knTM} and \eqref{kmTE} are rewritten as follows
\begin{align} \label{knTM}
k_{n} &= \frac{2 \pi n}{p_{\text{x}}} + k_{\text{t}} \\
\label{kmTE} k_{m} &= \frac{2 \pi m}{p_{\text{y}}}  \,
\end{align}
since the transverse component of the wavevector is leading towards $\hat{\mathbf{x}}$, as mentioned above.

Furthermore, by applying the same methodology we achieve the following expressions for the factors multiplying the TE/TM admittances in $Y_{\text{eq}}^{\text{I}}$
\begin{align} \label{NTE1_R}
\alpha_{nm}^{\text{TE}, (\text{in})} =& \bigg[ \frac{k_{n}}{k_{nm}} \frac{\cos(k_{n} w_{\text{x}}/2)}{\cos(k_{\text{t}} w_{\text{x}}/2)}  
\frac{(k_{\text{t}} w_{\text{x}})^2 - \pi^2 }{(k_{n} w_{\text{x}})^2 - \pi^2} \frac{\sin(k_{m} w_{\text{y}}/2)}{k_{m} w_{\text{y}}/2}\bigg]^2 \\
\label{NTM1_R}
\alpha_{nm}^{\text{TM}, (\text{in})} &=  \frac{k_{m}^2}{k_{n}^2} \alpha_{nm}^{\text{TE}, (\text{in})} \\
\alpha_{10}^{\text{TE, (RWG)}} &= \frac{1}{8 \pi^2}\frac{p_{\text{x}} p_{\text{y}}}{w_{\text{x}} w_{\text{x}}}\bigg[\frac{(k_{\text{t}} w_{\text{x}})^2 - \pi^2}{\cos(k_{\text{t}} w_{\text{x}}/2)}\bigg]^{2}\,.
\end{align}
which contribute to redefine the admittances taking place on the equivalent circuit. The factor associated with the $\text{TE}_{10}$ mode, $\alpha_{1}$, keeps being the one in \eqref{alpha1_TM}.
\subsection{Rectangular Waveguide -- Hard Waveguide Discontinuity (II)}

In order to model discontinuity II, we assume the RWG as the input media, and the HWG with dimensions $p_{\text{x}} \times h_{1}$ as the output one. That is, the \emph{feeding port} is set on the RWG side. In Sec.~\ref{air_SWG}, it was assumed that the $\text{TE}_{10}$-mode was the dominant field inside the RWG (both in evanescent or propagative state). This approximation is valid for frequencies below and above the cutoff frequency of this mode, and it will also be applied in this section. Thus, the \emph{incident field} is now the $\text{TE}_{10}$ mode of the RWG.

The interaction of this mode with the discontinuity excites all the possible modal solutions in both waveguides. In the RWG region, the $\text{TE}_{10}$ mode is highly dominant thus the modal expansion is exactly the same as that in \eqref{E2} and \eqref{H2}. In the HWG, the excitable modal solutions \emph{under the incidence} of the $\text{TE}_{10}$ are those which respect the same symmetrical conditions imposed by this last mode. After some mathematical and physical analysis, it can be concluded that modal solutions with even orders $n$ and $m$ can only be excited at the discontinuity. The electric-field expansion in the HWG region therefore admits to be described as: 
\begin{multline}
\label{expansion_PPW_new} \mathbf{E}^{\text{(HWG)}}(x, y) = \frac{1}{\sqrt{p_{\text{x}} w_{\text{y}}}} E_{00}^{\text{(HWG)}} \hat{\mathbf{y}} \\ + \bigg[\displaystyle\sum_{\substack{ \forall n > 0 \\ \forall m \ge 0}}   \gamma_{nm}^{\text{(HWG)}} E_{nm}^{\text{TE, (HWG)}} k_{n}^{\text{(HWG)}} \\ \times \cos(k_{n}^{\text{(HWG)}} x) \cos(k_{m}^{\text{(HWG)}} y) \bigg] \hat{\mathbf{y}}  \\ - \bigg[\displaystyle\sum_{\substack{\forall n \ge 0 \\ \forall m > 0}}  \gamma_{nm}^{\text{(HWG)}} E_{nm}^{\text{TM, (HWG)}} k_{m}^{\text{(HWG)}} \\ \times \cos(k_{n}^{\text{(HWG)}}x) \cos(k_{m}^{\text{(HWG)}} y)\bigg] \hat{\mathbf{y}} 
\end{multline}
for $n, m$ even numbers, and with 
\begin{align}
\gamma_{n0}^{\text{(HWG)}} &= \frac{2}{\sqrt{2}} \frac{1}{k_{n}^{\text{(HWG)}}} \frac{1}{\sqrt{p_{\text{x}} w_{\text{y}}}}\\
\gamma_{0m}^{\text{(HWG)}} &= \frac{2}{\sqrt{2}} \frac{1}{k_{m}^{\text{(HWG)}}} \frac{1}{\sqrt{p_{\text{x}} w_{\text{y}}}}\\
\gamma_{nm}^{\text{(HWG)}} &= \frac{1}{2} \frac{1}{k_{nm}^{\text{(HWG)}}} \frac{1}{\sqrt{p_{\text{x}} w_{\text{y}}}}\,. \\
k_{n}^{\text{(HWG)}} &= \frac{n \pi} {p_{\text{x}}}  \hspace{5 mm} \forall n \hspace{2 mm} \text{even} \\
k_{m}^{\text{(HWG)}} &= \frac{m \pi} {h_{1}}  \hspace{5 mm} \forall m \hspace{2 mm} \text{even} 
\end{align}

The field profile at the discontinuity is that used in Sect.~\ref{air_SWG}, specified in \eqref{profile_TE}. Taking this into account, and imposing the continuity equations in \eqref{EQ1}, \eqref{EQ2} and \eqref{EQ3}, we obtain expression of an equivalent admittance for discontinuity II which includes the effect of all the higher order harmonics in the HWG:
\begin{multline}
Y_{\text{eq}}^{\text{II}} = \displaystyle\sum_{\substack{\forall n > 0 \\ \forall m \ge 0}}    \alpha_{nm}^{\text{TE, (HWG)}} Y_{nm}^{\text{TE, (HWG)}} \\ + \displaystyle\sum_{\substack{\forall n \ge 0 \\ \forall m > 0}} \alpha_{nm}^{\text{TM, (HWG)}} Y_{nm}^{\text{TM, (\text{HWG})}} 
\end{multline}
where the factors $\alpha$ can be defined as follows 
\begin{equation}
\alpha_{1}\alpha_{n0}^{\text{TE, HWG}} =  32 \pi^2 \frac{w_{\text{x}} w_{\text{y}}}{p_{\text{x}} h_{1}} \bigg[\frac{\cos(k_{n}^{\text{(HWG)}} w_{\text{x}}/2)}{(k_{n}^{\text{(HWG)}} w_{\text{x}})^2 - \pi^2} \bigg]^2 
\end{equation}
\begin{multline}
\alpha_{1} \alpha_{nm}^{\text{TE, (HWG)}} = 64 \pi^2  \frac{w_{\text{x}} w_{\text{y}}}{p_{\text{x}}h_{1}} \times \\ \bigg[\frac{k_{n}^{\text{(HWG)}}}{k_{nm}^{\text{(HWG)}}} \frac{\cos(k_{n}^{\text{(HWG)}} w_{\text{x}}/2)}{(k_{n} ^{\text{(HWG)}} w_{\text{x}})^2 - \pi^2} \frac{\sin(k_{m}^{\text{(HWG)}} w_{\text{y}}/2)}{k_{m}^{\text{(HWG)}}w_{\text{y}}/2} \bigg]^2 
\end{multline}
\begin{equation}
\alpha_{1}\alpha_{0m}^{\text{TM, (HWG)}}  =  \frac{32}{\pi^2} \frac{w_{\text{x}} w_{\text{y}}}{p_{\text{x}} h_{1}} \bigg[\frac{\sin(k_{m}^{\text{(HWG)}} w_{\text{y}}/2)}{k_{m} ^{\text{(HWG)}}w_{\text{y}}/2}\bigg]^2
\end{equation}
\begin{equation}
\alpha_{nm}^{\text{TM, (HWG)}} = \bigg[\frac{k_{m}^{\text{(HWG)}}}{k_{nm}^{\text{(HWG)}}} \bigg]^2 \alpha_{nm}^{\text{TE, (HWG)}} \,.
\end{equation}
The parameter $\alpha_{1}$ is the one defined in \eqref{alpha1_TM}.
The admittance $Y_{\text{eq}}^{\text{II}}$ is connected to the input and output transmission lines in grey and orange colors in \Fig{fig1}. They are the input and output lines of this discontinuity, corresponding to the $\text{TE}_{10}$ and TEM modes respectively. The characteristic admittances are given by 
$\alpha_{1} Y_{10}^{\text{TE, (RWG)}}$ and $\alpha_{2}Y_{\text{TEM}}^{(\text{HWG1})}$
with
\begin{equation}
\alpha_{1} \alpha_{2} =  \frac{16}{\pi^2} \frac{w_{\text{x}} w_{\text{y}}}{p_{\text{x}} h_{1}}\,.
\end{equation}
and $Y_{\text{TEM}}^{(\text{HWG1})} = 1/\eta_{0}$.
\subsection{Hard Waveguide -- Hard Waveguide Discontinuity (III)}

In order to model this kind of discontinuity, we take into account the previous conclusions extracted from the rest of discontinuity types. Now, two different HWGs are in contact at the discontinuity. The input HWG, being the output region in the previous section, has dimensions $p_{\text{x}} \times h_{1}$. The output HWG has dimensions $p_{\text{x}} \times h_{2}$. The feeding mode is now the TEM mode coming from the input HWG. At the discontinuity, the field at both sides is expressed as in \eqref{expansion_PPW_new}, but employing the correct dimensions ($p_{\text{x}}$ and $h_{1}$ for the input region; and $p_{\text{x}}$ and $h_{2}$ for the output region) in the wavenumbers $k_{n}$ and $k_{m}$.

The field profile at the discontinuity is proportional to the TEM-mode in the lowest-height HWG:
\begin{equation}
\mathbf{E}_{\text{d}} = A \hat{\mathbf{y}}
\end{equation}
Assuming $h_{1} < h_{2}$, and applying again the boundary conditions in \eqref{EQ1}, \eqref{EQ2} and \eqref{EQ3} we obtain:
\begin{equation}\label{transformers_HWG_HWG}
E_{nm}^{\text{TE/TM, (HWG1/HWG2)}} = 0 \hspace{5 mm} \text{if} \hspace{1 mm} n \ne0 \hspace{1 mm} \text{or} \hspace{1 mm} m \ne 0 \,,
\end{equation}
indicating that no higher-order modes are considered under this approximation. This approximation works well up to the excitation any of the first higher-order mode.
Therefore, the fundamental modes (TEM) of each of the waveguides are represented by transmission lines with characteristic admittances $Y_{\text{TEM}}^{(\text{HWG1})} = Y_{\text{TEM}}^{(\text{HWG2})} = 1/\eta_{0}$, being those in orange and yellow in \Fig{fig1} respectively. The parameter $\alpha_{3}$ is defined as follows:
\begin{equation}\label{alpha3}
\alpha_{1} \alpha_{2} \alpha_{3} = \frac{h_{1}}{h_{2}}\,.
\end{equation}
It is worth remarking that, in case that $h_{1} > h_{2}$, equation \eqref{alpha3} becomes:
\begin{equation}
\alpha_{1} \alpha_{2} \alpha_{3} = \frac{h_{2}}{h_{1}}\,.
\end{equation}
In a first approach, the shunt admittance connecting both transmission lines can be neglected, $Y_\mathrm{eq}^{\text{III}} = 0$, since it is formally formed by all the higher modes $E_{nm}^{\text{TE/TM, (HWG1) / (HWG2)}}$, non-excited under the conditions imposed at the discontinuity according to \eqref{transformers_HWG_HWG}. This constitutes an approximation, expected to work well up to the excitation of the higher-order modes in any of the HWGs.

\subsection{Numerical Examples: Reflective Structures}

\begin{figure}[!t]
	\centering
	\subfigure[]{\includegraphics[width= 0.95\columnwidth]{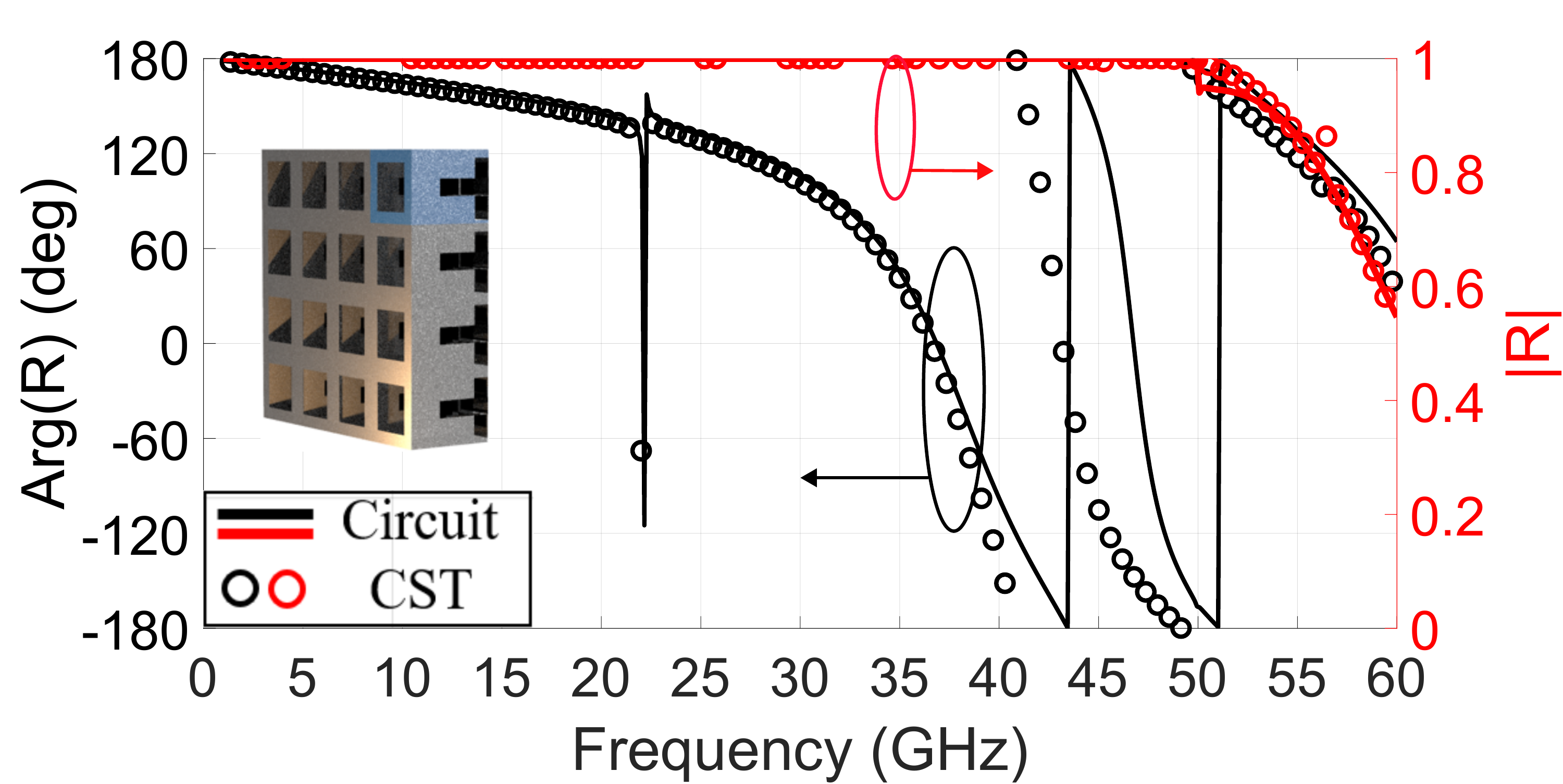}}
     \subfigure[]{\includegraphics[width= 0.515\columnwidth]{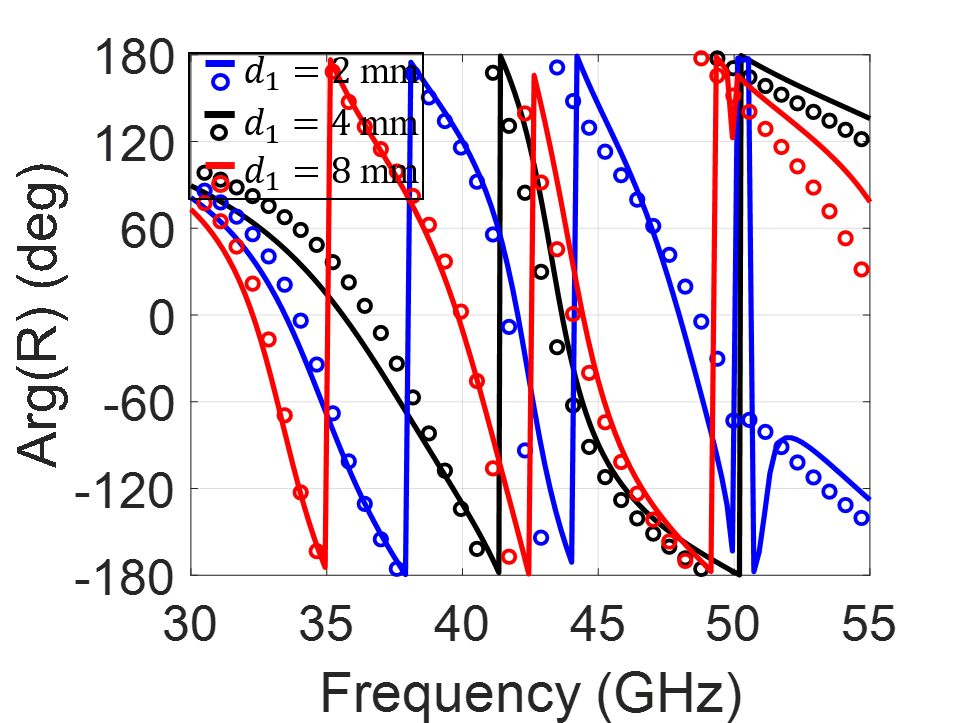}}
     \hspace{-0.5cm}
     \subfigure[]{\includegraphics[width= 0.515\columnwidth]{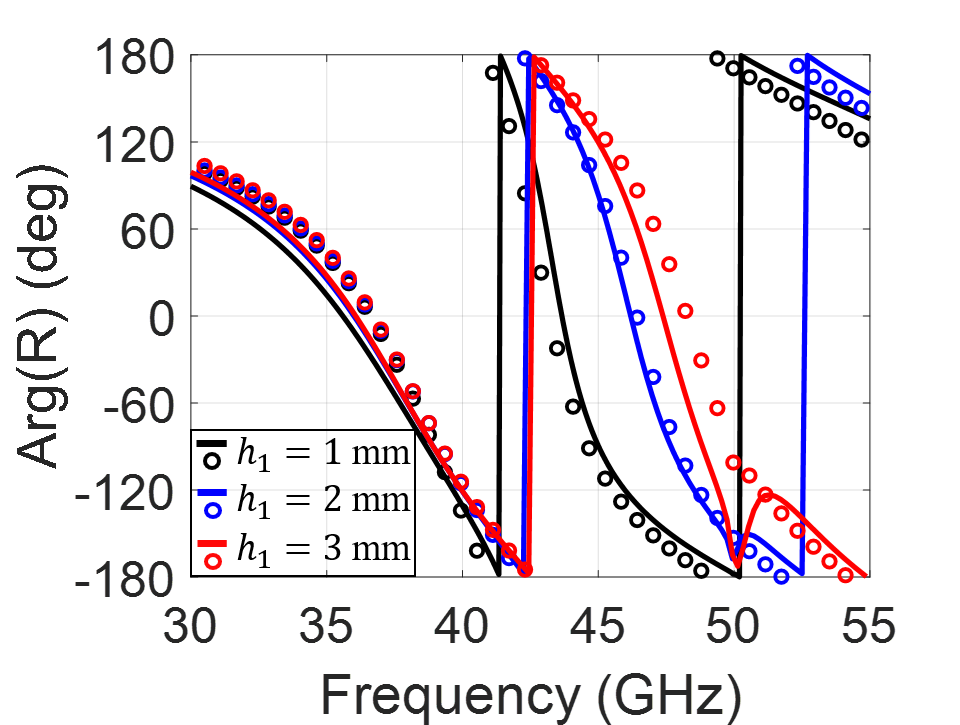}}
	\subfigure[]{\includegraphics[width= 1\columnwidth]{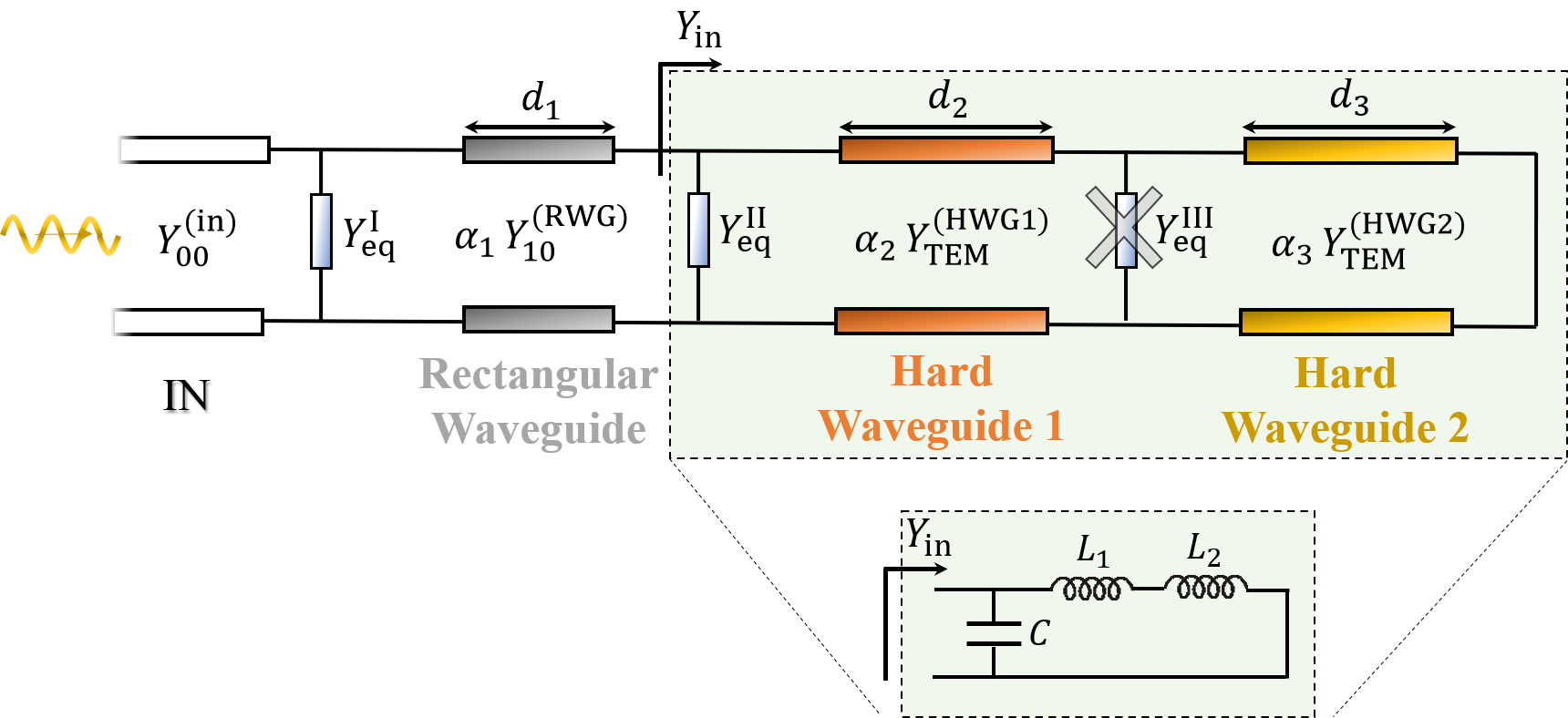}}
	\caption{Reflection coefficient $R$ of a reflective 3D structure for $d_1=4\,$mm and $h_1=1\,$mm. (a) Phase and amplitude. (b) Effect of varying $d_1$. (c) Effect of varying $h_1$. In all cases, colored solid lines and circles correspond to the analytical and full-wave results, respectively. (d)  Analytical circuit model. Geometrical parameters: $p_{\text{x}}=p_{\text{y}}=6\,$ mm, $w_{\text{x}} = w_{\text{y}} = 5\,$mm, \linebreak $d_2=4\,$mm, $d_3=0.5\,$mm, and $h_2=3\,$mm. Normal incidence. }
	\label{fig2}
\end{figure}

Now, we present some numerical results to validate and test the capabilities of the analytical equivalent circuit. We initially consider the 3D metamaterial structure included as an inset in \Fig{fig2}(a), with its associated circuit shown in \Fig{fig2}(d). It is a fully-metallic device, short circuited at its end, that operates as a reflective structure. 
From the circuit point of view, the input admittance seen from discontinuity III through the short-circuited line can be represented as
\begin{equation}
Y_{\text{in}}^{(\text{HWG2})} = -\text{j} \alpha_{3} Y_{\text{TEM}}^{(\text{HWG2})} \cot(k_{0} d_{3})\,.
\end{equation}
The longitudinal cross-shaped slots and the associated waveguide stretches allow to control the phase of the reflected wave in an efficient manner. 

\Fig{fig2}(a) illustrates a comparison  between results obtained by our approach (solid lines) and the full-wave simulator (circles) CST Microwave Studio. The parameters computed are the phase (black) and amplitude (red) of the reflection coefficient $R$. As observed, there is a good agreement between analytical and full-wave results in a wide range of frequencies. For the dimensions of the unit cell ($6\times6\,\text{mm}^2$), the onset frequency of the first higher-order harmonic is $50$ GHz for normal incidence. Naturally, the computation time for the analytical circuit (order of seconds) is much less than that of commercial software CST (order of minutes or hours could be needed). For 1001 frequency points, the circuit approach took less than 2 seconds in giving the solution while CST took more than 400 seconds in the simplest case.

The square-waveguide section is under cutoff below 30 GHz for the selected dimensions ($5\times5\, \text{mm}^2$). This implies that only evanescent modes are excited inside the 3D metagrating below 30 GHz. Nonetheless, in cases where the RWG length is short enough, modes of evanescent nature in the RWG (e.g. TE$_{10}^{(\mathrm{RWG})}$) may couple to the HWG section. This is due to the \emph{slow} decay rate of the amplitude of the $\text{TE}_{10}$ mode along the short RWG section. The HWG is therefore excited by the evanescent $\text{TE}_{10}$ mode, manifested by the propagation of its fundamental mode, now identified as a propagating TEM mode. This excitation via evanescent waveguides is analogous to the well-known tunnel effect in quantum physics \cite{Griffiths95}, and is particularly relevant to design transmitting structures via opaque waveguides. Such is the case described in Sec~\ref{transmitting}.

The spectral evolution of the reflection coefficient plotted in \Fig{fig2}(a) exhibits a rich phenomenology, specially for frequencies higher than 30 GHz. However, a sudden resonance jump appears around $22\,$GHz, identified as a sudden $\lambda/2$-resonance given in the HWG regions. The total length of the two HWG sections (transmission-line sections) sums up $D = d_{2} + d_{3} = 4.5\,\text{mm}$. The corresponding $\lambda/2$ resonance would be expected at $33.3\,$GHz in a case where the HWGs were isolated. However the HWGs sections are coupled to a RWG section, whose common discontinuity junction is modeled by the admittance $Y_{\text{eq}}^{\text{II}}$, predominantly capacitive. The presence of this capacitance is the responsible of the resonance shifting to lower frequencies. In this case down to $22\,$GHz. The rest of RWG section contributes as an inductive load though its influence over the rest of the structure is not quite relevant. The phase evolution starts to vary faster in frequency beyond $30\,$GHz, frequency at which the $\text{TE}_{10}$-mode becomes propagative. Both the $\lambda/4$ peak as well as this fast evolution of the phase at higher frequencies is well caught by the equivalent circuit, since it predicts all this phenomenology. 

The red curve and circles show the amplitude evolution of the reflection coefficient in \Fig{fig2}(a). As expected, the reflective character of the cell invokes full reflection up to 50 GHz. Above 50 GHz, the first higher-order harmonic in the air region becomes propagative, thus the reflected power is split in two: part is carried by the incident wave and part by the higher-order harmonic. This is the reason why the reflection coefficient represented in the figure is no longer one beyond $50\,$GHz. Again, this phenomenon is well captured by the equivalent circuit.

\Figs{fig2}(b) and (c) illustrate the spectral evolution of the reflection phase when some geometrical parameters of the structure are modified. As expected, the increase of the lengths of the RWG and HWG sections has a direct impact on the phase response of the system. \Fig{fig2}(b) shows the phase modification when the length of the RWG section $d_{1}$ is varied. A comparison is made between the analytical-circuit results (solid lines) and CST (circles), showing a very good agreement in all cases. Slight differences can be appreciated at high frequencies, due to the complexity of the structure. Moreover, the model seems to be more accurate for higher values of $d_{1}$. For $d_{1} = 2\,$mm, differences between CST and the circuit approach can be appreciated at lower frequencies in comparison with cases assuming higher values of $d_{1}$. The reason behind this disagreement is related to the lack of information associated with higher-order modes inside the RWG. The model assumes the excitation of the $\text{TE}_{10}$-mode (both in evanescent and propagative nature), and does avoid the excitation of the rest of higher-order modes. When $d_{1}$ is a large value, and the $\text{TE}_{10}$-mode is evanescent (under $30\,$GHz), its amplitude decays along the RWG and does not reach the end of the RWG. As $d_{1}$ decreases, this amplitude may not decay along the RWG length and the modal field can reach the end of the waveguide, coupling to the next waveguide (HWG). For $d_{1} \ll \lambda$, the $\text{TE}_{10}$-mode does arrive to the end, but also some of the higher-order modes that have not been taken into account. When the frequency is higher than $30\,$GHz, the $\text{TE}_{10}$-mode becomes propagative and its amplitude stops decaying. The model captures well this fact. However the presence of the rest of higher-order modes is not considered in the circuit model, thus a lack of accuracy is expected when $d_{1} \ll \lambda$. As it can be appreciated in Fig. 2(b), this disagreement from $30\,$GHz is not visible for $d_{1} = 4$ and $d_{1} = 8\,$mm.  

\Fig{fig2}(c) shows the phase variation when $h_{1}$ is modified. The lower the value of $h_1$ is, the greater  the observed phase shift is. Physical insight into this phenomenon can be achieved by means of the analytical circuit model. For HWG with short lengths ($d_2, d_3 \ll \lambda$), the short-circuited HWG sections contribute to the circuit model with a pure inductive term $L = L_1 + L_2$. In addition, the equivalent admittance that models the RWG-HWG discontinuity, $Y_\mathrm{eq}^\mathrm{II}$, contributes with a pure capacitive term $C$ at frequencies above the RWG cutoff. Thus, the RWG-HWG-short section, whose input admittance is $Y_\mathrm{in}$ [see \Fig{fig2}(d)], can be simply described as a $LC$ tank. The fact of decreasing the height of the HWG provokes that its lower and upper metallic plates are now closer. As a consequence, the associated capacitance $C$ increases, shifting down in frequency the position of the the curves in \Fig{fig2}(c). Note that this effect is not appreciable near the cutoff of the square waveguide  (30 GHz for the selected geometry), as the capacitive term $C$ is not dominant yet. 

\begin{figure}[t!]
	\centering
     \subfigure[]{\includegraphics[width= 1.05\columnwidth]{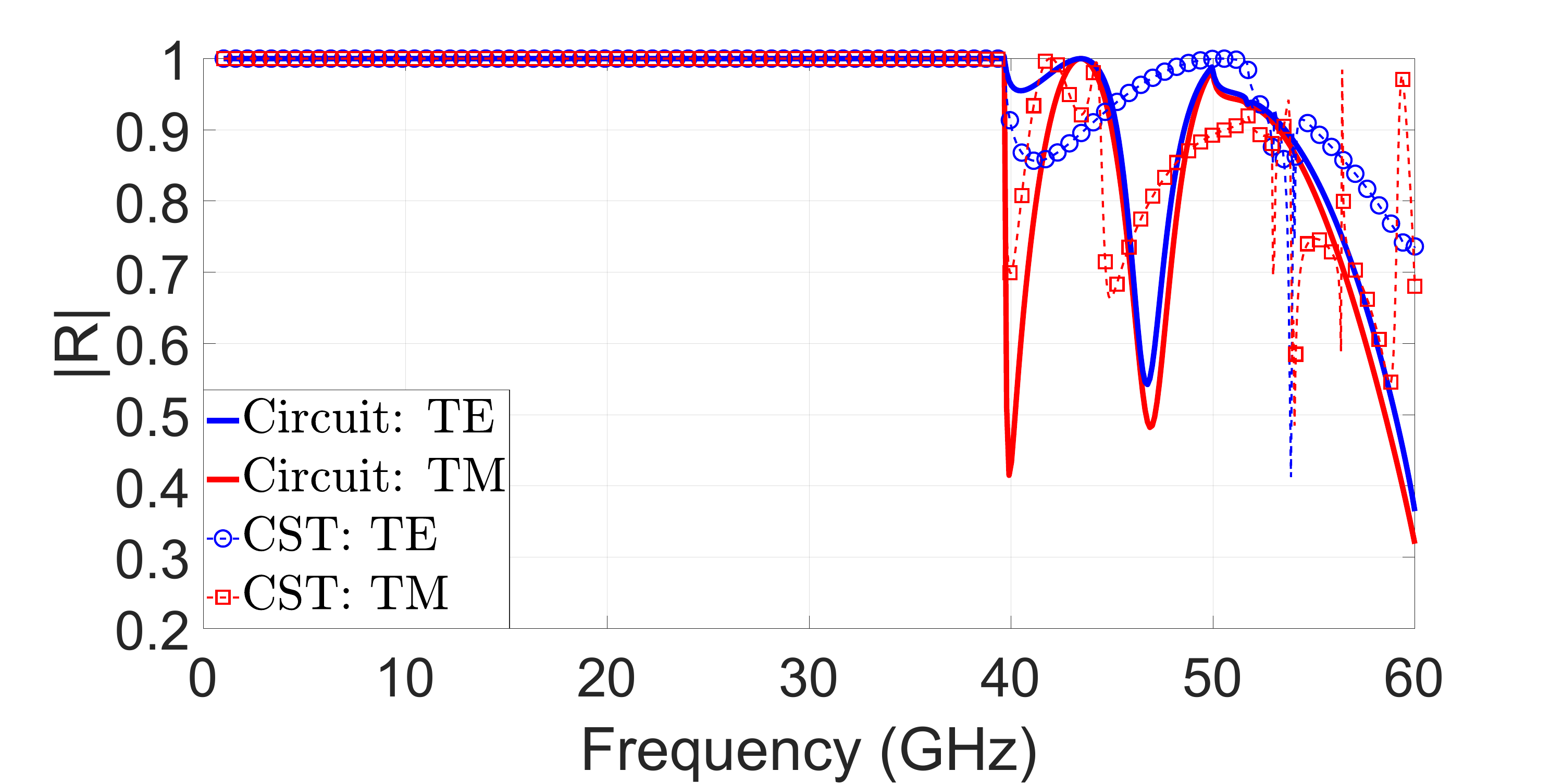}}
	\subfigure[]{\includegraphics[width= 1.05\columnwidth]{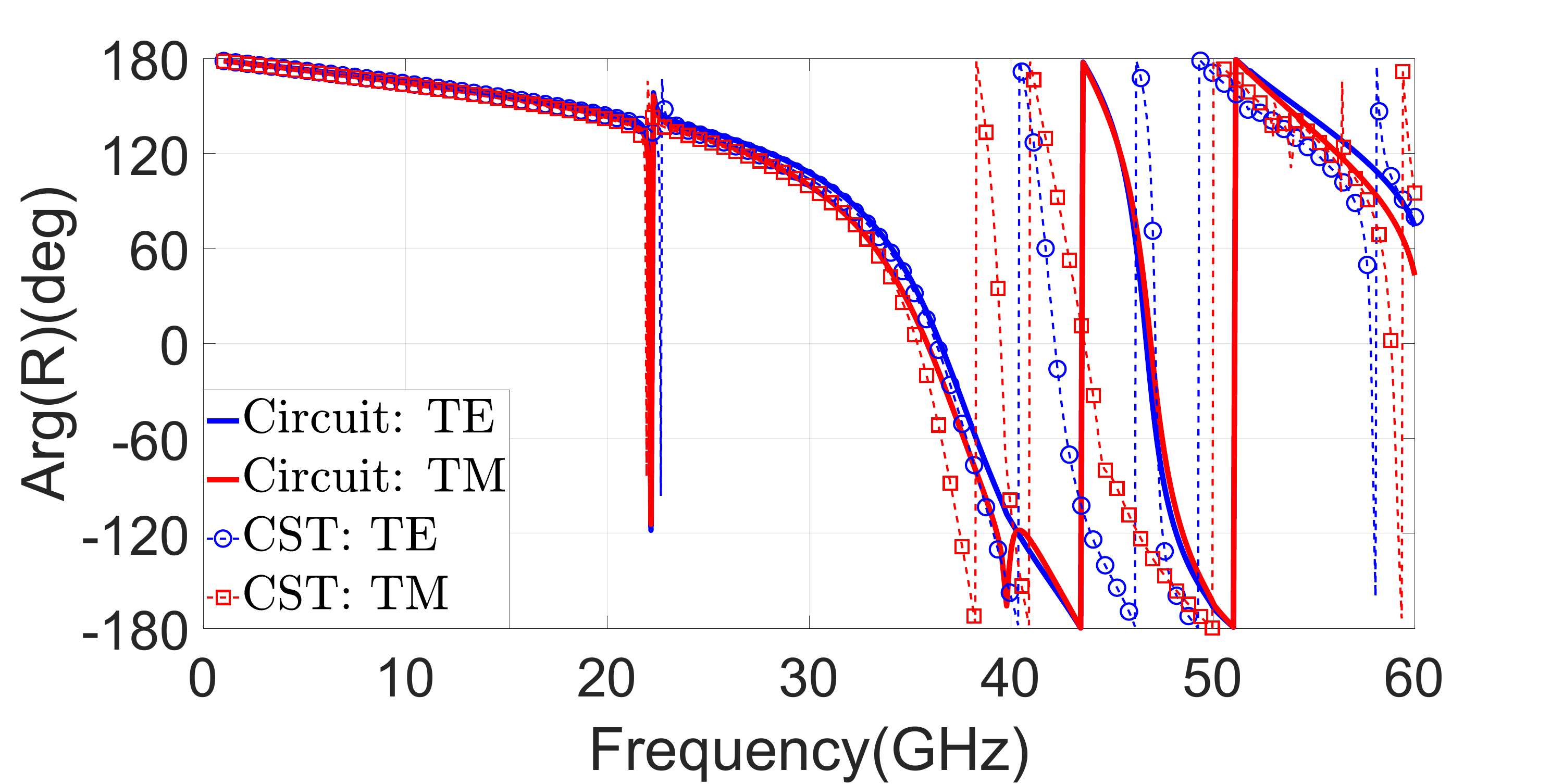}}
	\caption{Reflection coefficient $R$ of a reflective 3D structure under oblique incidence conditions \linebreak ($\theta=15$ deg). (a) Amplitude. (b) Phase. Geometrical parameters: $p_{\text{x}}=p_{\text{y}}=6\,$mm, $w_{\text{x}} = w_{\text{y}} = 5\,$mm, \linebreak $d_1=d_2=4\,$mm, $d_3=0.5\,$mm, $h_1=1\,$mm and $h_2=3\,$mm.}
	\label{Oblicua}
\end{figure}

Finally, results considering oblique incidence have been plotted in \Fig{Oblicua}. TE and TM incidences have been included, manifested by the incidence of a plane wave with incidence angle \linebreak $\theta = 15$ deg.  The results provided by the circuit model fit well with those obtained by CST up to 40 GHz approximately. Beyond this frequency the agreement notably deteriorates, due to the excitation of additional modes that are no included in the mathematical circuit derivation. Specially, higher-order modes inside the RWG are now determinant, as the $\text{TE}_{11}$, with cutoff frequency $f_{11} = 42.43\,$GHz. Since this mode is excluded from the circuit derivation, it is expected the degradation manifested in \Fig{Oblicua} near and beyond $f_{11}$. Though it constitutes an important limitation of our approach, the model is still wideband, exhibiting good performance for cell dimensions larger than $\lambda/2$. It is worth remarking that all the physical phenomena below $f_{11}$ are well caught, as the resonance in the reflection phase around $22\,$GHz.

\subsection{Numerical Examples: Transmitting Structures}\label{transmitting}

The proposed 3D metastructure can operate in reflection and transmission modes. By minor modifications of the geometry of the waveguides that form the 3D metastructure, a design initially proposed for reflection operation can be converted into one with transmission operation. From a practical point of view, this is an interesting feature of the proposed 3D structures. When we normally consider other types of structures, specific designs are made for reflection and transmission, which can sometimes be very different from each other. In transmission mode, the 3D device can control both  amplitude and phase of the scattered waves by properly adjusting the geometrical parameters of rectangular and hard waveguides. 

\begin{figure}[!t]
	\centering
	\subfigure[]{\includegraphics[width= 1.015\columnwidth]{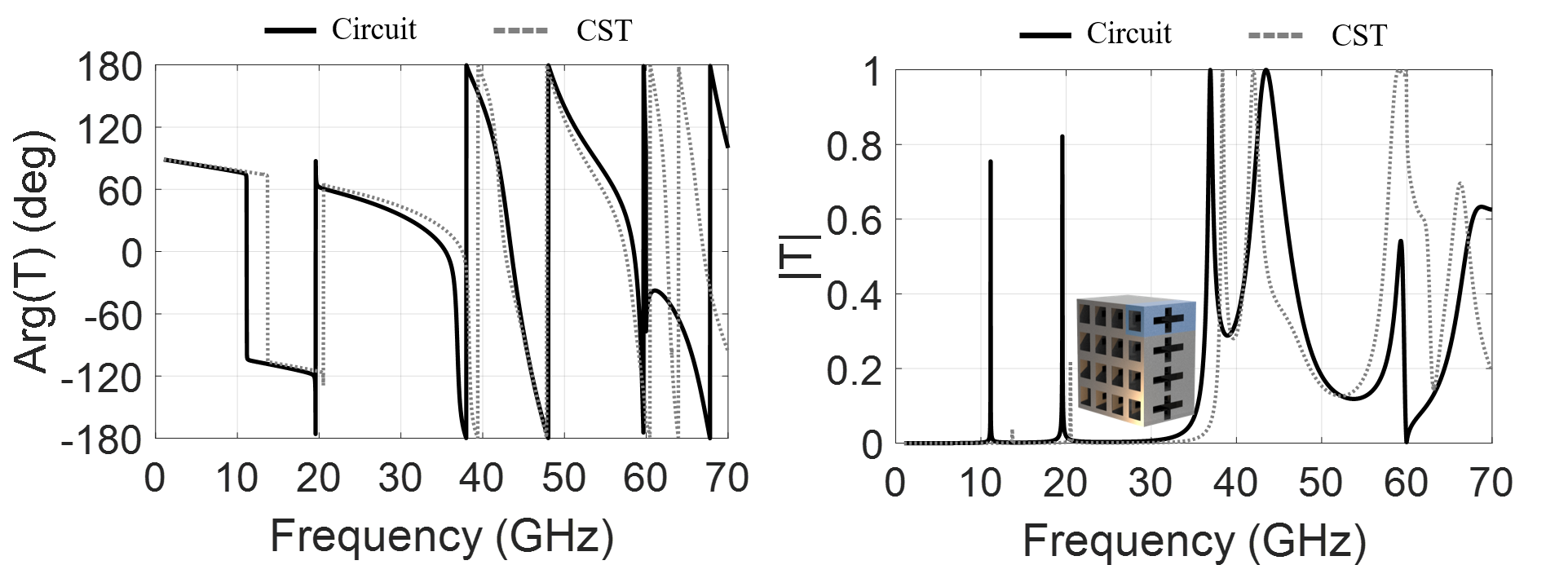}} \\
     \subfigure[]{\includegraphics[width= 0.515\columnwidth]{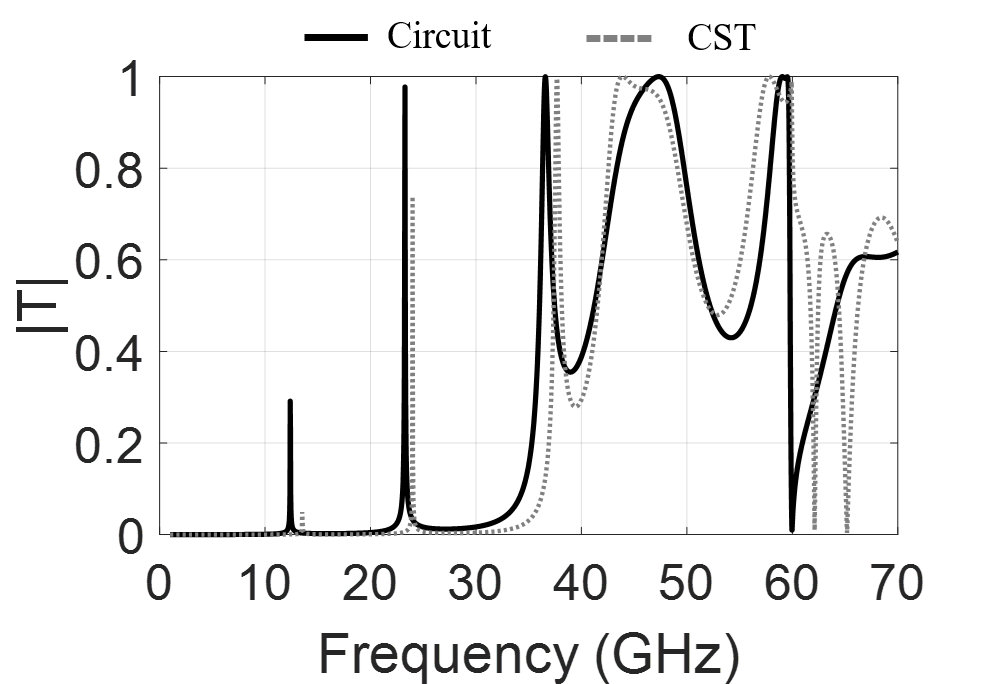}}
     \hspace*{-0.5cm}
    \subfigure[]{\includegraphics[width= 0.515\columnwidth]{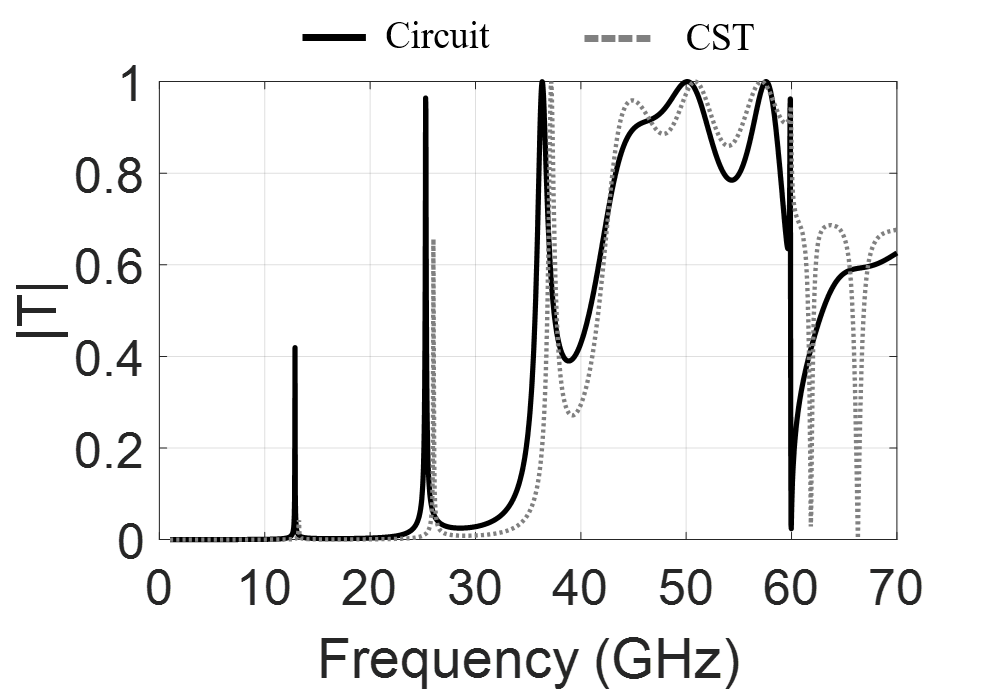}}
	\caption{(a) Transmission coefficient $T$ (phase and amplitude) of a transmitting 3D structure ($h_1=1\,$mm). \linebreak (b) $|T|$ for $h_1 = 2\,$mm. (c) $|T|$ for $h_1 = 3\,$mm. The analytical circuit model corresponds to that shown in \Fig{fig1}(c). Geometrical parameters: $p_{\text{x}}=p_{\text{y}}=5\,$mm, $w_{\text{x}} = w_{\text{y}} = 4\,$mm,  $d_1 = 4\,$mm, $d_2=4\,$mm, $d_3=1\,$mm, $d_4=4\,$mm, $d_5=4\,$mm, and $h_2=3\,$mm. Normal incidence is assumed.}
	\label{fig3}
\end{figure}

\Fig{fig3} shows the transmission coefficient $T$ (phase and amplitude) in the 3D transmitting structure. The performance of the analytical equivalent circuit is tested against full-wave results extracted from CST. A good agreement is generally observed in a wideband range. For the calculation of 5001 frequency points, the circuit approach lasted less than 4 seconds while CST took more than 10 minutes. Time-domain full-wave solvers such as CST are highly dependent on the size of the structure under analysis. Transmitting structures are longer than reflective structures, so computation times notably increase. Remarkably, our circuit approach is, for all practical purposes, independent from the considered size. In fact, even-odd excitation techniques \cite{Torres2016, Molero2016} can be considered for the analysis of longitudinally-symmetric transmitting structures such as the one considered here (symmetric with respect to the plane $D=d_1+d_2/2$), approximately halving the computation time. 

In \Fig{fig3}, two resonant transmission peaks appear  below the cutoff frequency of the RWG (37.5 GHz for the selected geometry). As previously discussed,   evanescent waves travel along the RWG and, at some particular frequencies, may couple to the HWG sections led by the fundamental TEM mode. Thus, narrowband transmission does occur below the cutoff of the RWG. The two narrowband transmission peaks are caused by $\lambda/2$ and $\lambda$ resonances in the HWG sections. Ideally, for the selected length of the HWG sections (9 mm in total), the $\lambda/2$ resonance would be located at 16.67 GHz if the HWG sections were isolated from the RWG and uniform in height. In practice, \Figs{fig3}(a), (b), (c) show that the $\lambda/2$ resonance is located at 11.2 GHz, 12.4 GHz and 13 GHz, respectively, for an increasing height $h_1$. As the height of the HWG increases, the simulated resonance approaches the theoretical estimation, as the capacitive contribution of $Y_\mathrm{eq}^\mathrm{II}$ and $Y_\mathrm{eq}^\mathrm{V}$ (RWG--HWG transitions) is less prominent. Same rationale holds for the $\lambda$ resonance created by the HWG.

The circuit approach stops being accurate beyond 60 GHz. This coincides with the onset of the first higher harmonic in the HWG, at 60 GHz (larger dimension/periodicity $p_{\text{x}} = 5\,$mm). In any case, the accuracy of the model is quite good below this frequency, covering an actually large bandwidth.

\subsection{Independent Polarization Control}

\begin{figure}[t!]
	\centering
	\subfigure{\includegraphics[width= 1.1\columnwidth]{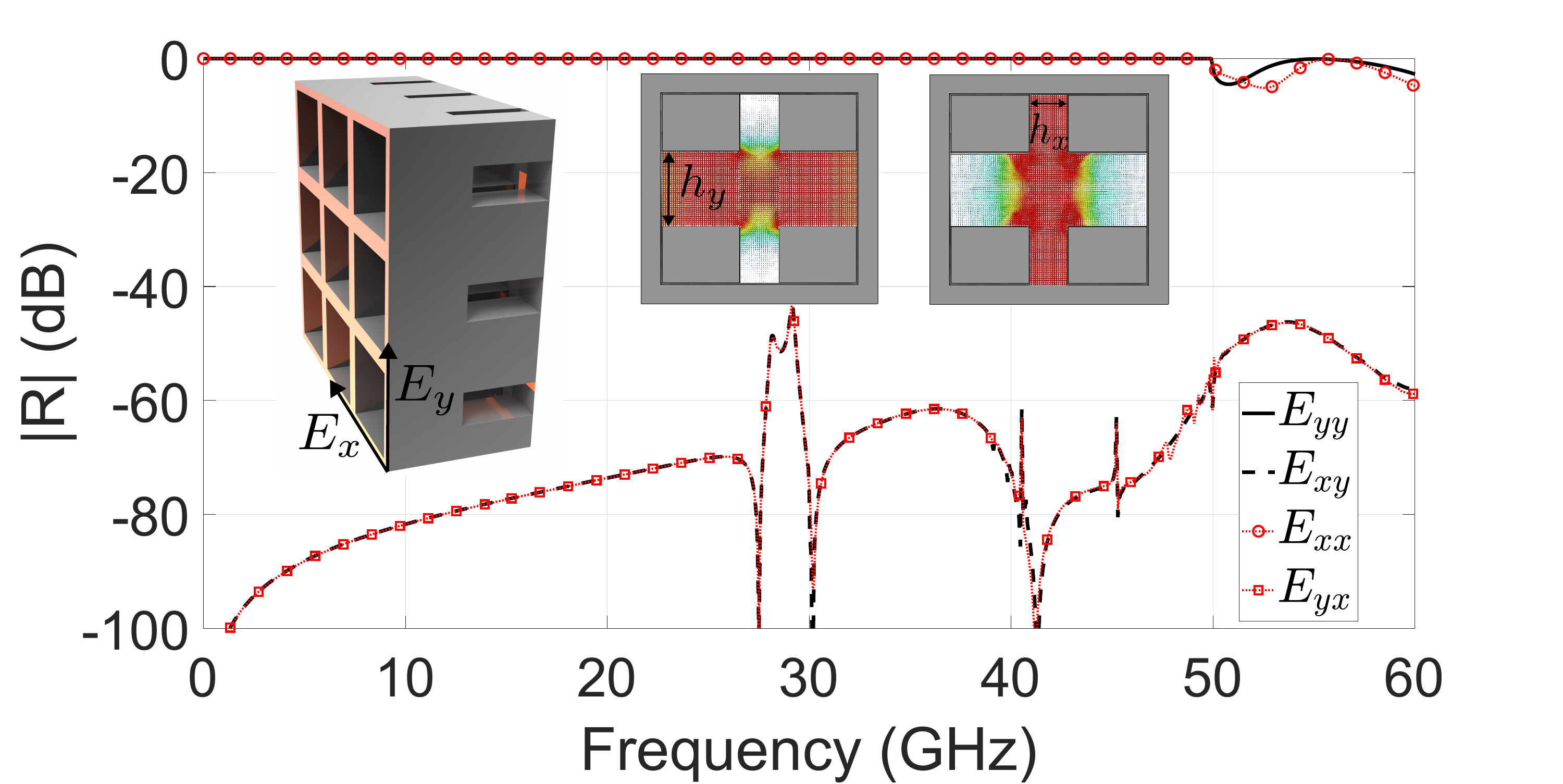}}
	\caption{Amplitude of the reflection coefficient of a reflective 3D structure. Co-pol ($E_{xx}, E_{yy}$) and cross-pol ($E_{xy}, E_{yx}$) terms are plotted. Geometrical parameters: $p_{\text{x}} = p_{\text{y}}=6\,$mm, \linebreak $w_{\text{x}} = w_{\text{y}} = 5\,$mm, $d_1=4\,$mm, $d_2=6\,$mm, $h_{\text{x}}=1\,$mm, $h_{\text{y}}=2\,$mm.}
	\label{PPWpartido}
\end{figure}

Another interesting property of the proposed 3D metamaterial is the independent polarization control of its two orthogonal linear states (X and Y polarizations). This is a key feature that can be exploited for the efficient design of polarizer devices. Independent polarization control is possible thanks to the use of the longitudinal direction as an additional degree of freedom. As was already shown in \cite{Carlos_MTT3D, Molero3D_2020}, resonators perforated on the waveguide walls in the XZ-plane controls X polarization and are opaque for Y polarization. The opposite situation is achieved when the resonators are perforated on the YZ wall. It is true that specific 2D metasurface designs may achieve a certain degree of polarization independence \cite{Mercader2021, Omar2018}. However this independence is highly sensitive to the scatterer shape and geometry. In 3D architectures, the resonator shapes do not break the polarization-independence property. Focusing on the structure under consideration in the present paper, independent polarization control can be easily achieved by placing an additional perpendicular HWG section. Thus, the HWG section in the 3D structure is now stretched along the $x$ and $y$ directions, as the insets in \Fig{PPWpartido} illustrate.

In \Fig{PPWpartido}, the orthogonal polarization independence is tested in a reflective structure (short-circuited at its end). \Fig{PPWpartido} shows the amplitude of the reflection coefficient $R$ for two perpendicular polarizations  at normal incidence. In the legend, $E_{ij}$ ($i,j = \{x,y\}$) states for the electric field associated to input $i-$polarized (horizontal)  and output  $j$-polarized  (vertical) waves. Thus, subindexes $ii$ and $jj$ represent the co-polarization (co-pol) terms while $ij$ and $ji$ represent the cross-polarization (cross-pol) terms. The cross-pol level is under $-40$ dB in a wideband range. Therefore, a great polarization independence can be claimed for the horizontal and vertical polarizations. Note that, from 50 GHz onwards, the single-mode behaviour of the structure ceases as the second mode of the HWG becomes propagative, as well as the first-orders harmonics in the air region.

For a visual representation, the inset in \Fig{PPWpartido} depicts the electric field profile of the fundamental mode for the horizontal and vertical polarizations. The electric field profile has been extracted with full-wave simulations in the commercial software. When the incident wave is polarized along the $y$ axis, the entire electric field is confined within the horizontal HWG. Cutting the structure at the center part does not significantly affect the electric-field pattern, which still shows a TEM-like profile. This fact leads to the low cross-pol coupling level evidenced in \Fig{PPWpartido}. The scenario is identical for a $x$-polarized incident wave. Most of the electric field is confined in the vertical HWG, also leading to a TEM-like profile. Polarization independence can be advantageously used for the design of polarizer devices, by simply tuning the length of the vertical and horizontal slots. Thus, a $0^{\text{o}}$-phase or  $180^{\text{o}}$-phase resonances can be achieved when the length of the short-circuited slot is $\lambda/4$ or $\lambda/2$, respectively \cite{Balmaseda2023}. In addition, the polarization independence is also advantageous from the circuital point of view, since and individual and independent circuit approach can be employed for the control of each of the polarizations ($x$ and $y$). The global problem, involving both polarizations, is split in two independent subproblems (one for the $x$ component, one for the $y$ component). 

\section{Related 3D Structures}

In this section, we show that the proposed analytical circuit approach can be used not only for the analysis of the original 3D device shown in \Fig{fig1}, but also for the characterization of related 3D metastructures and metagratings. As it will be shown, under some circumstances, related 3D structures present similar electromagnetic behavior to the originally proposed 3D metamaterial, thus the analytical equivalent circuit is still applicable. 

\begin{figure}[t!]
	\centering
	\subfigure{\includegraphics[width= 0.7\columnwidth]{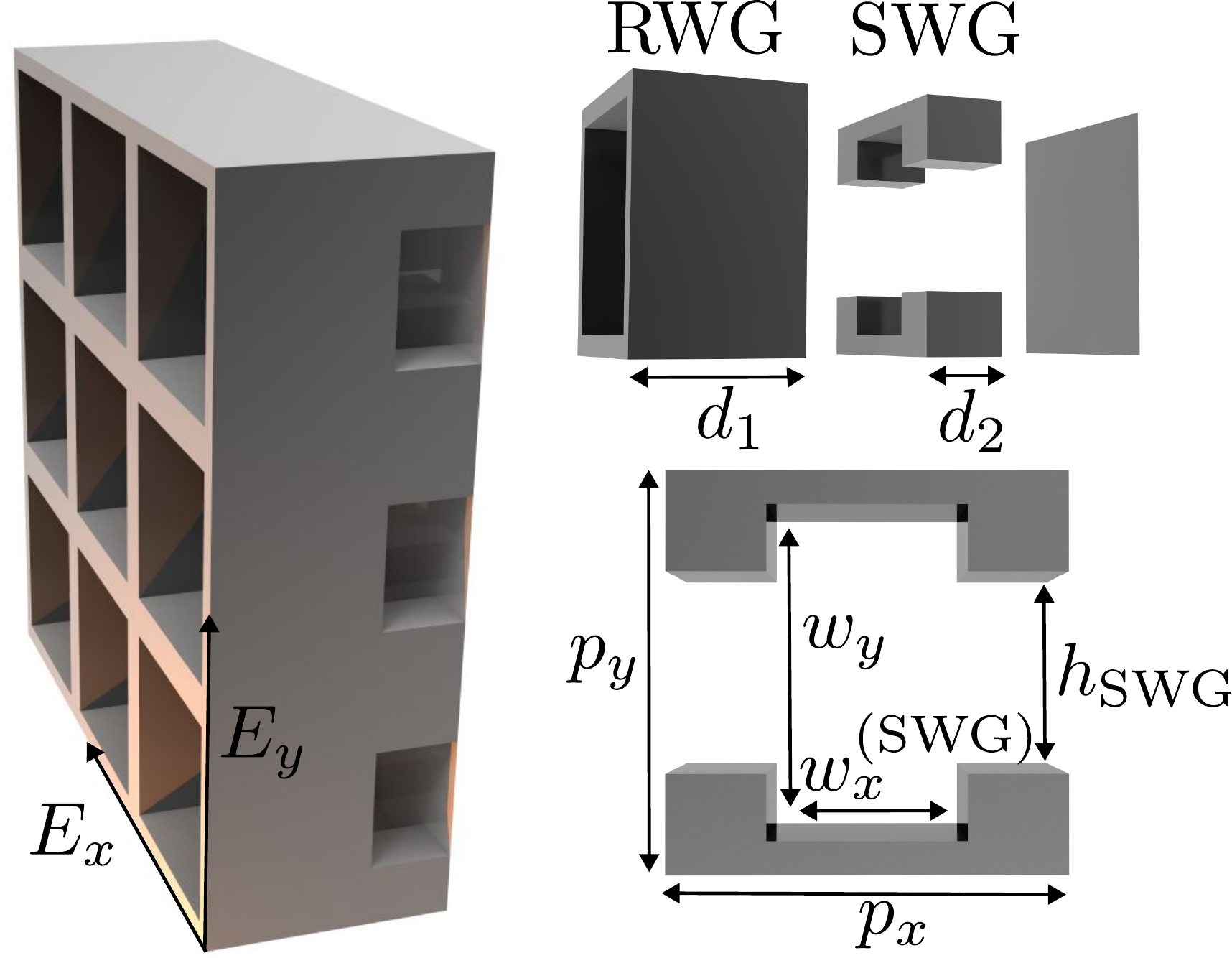}}
	\caption{``Alternative" 3D metamaterial formed by RWG and SWG sections. Note that its internal structure differs from the ``original" 3D metamaterial (RWG and HWG sections) shown in \Fig{fig1}(a). Geometrical parameters: \linebreak $p_{\text{x}}=p_{\text{y}}=6\,$ mm, $w_{\text{x}} = w_{\text{y}}= 5\,$mm, $w_{\text{x}}^{\text{(SWG)}} = 3\,$mm, $d_1=4\,$mm, $d_2=6\,$mm. The parameter  $w_{\text{y}}$ is identical for the RWG and SWG regions.} 
	\label{SWG}
\end{figure}

Specifically, we focus on the alternative version of the original 3D metamaterial shown in \Fig{SWG}. The alternative structure is similar in all aspects to the original one, except for the waveguide sections where the longitudinal slots are inserted. In the alternative 3D structure (\Fig{SWG}), the upper and lower metallic walls are not stretched along the $y$ direction in the slotted waveguide (SWG) region. This leads to a different field excitation in the SWG \cite{9696220}.

\begin{figure}[t!]
	\centering
     \subfigure[]{\includegraphics[width= 1.05\columnwidth]{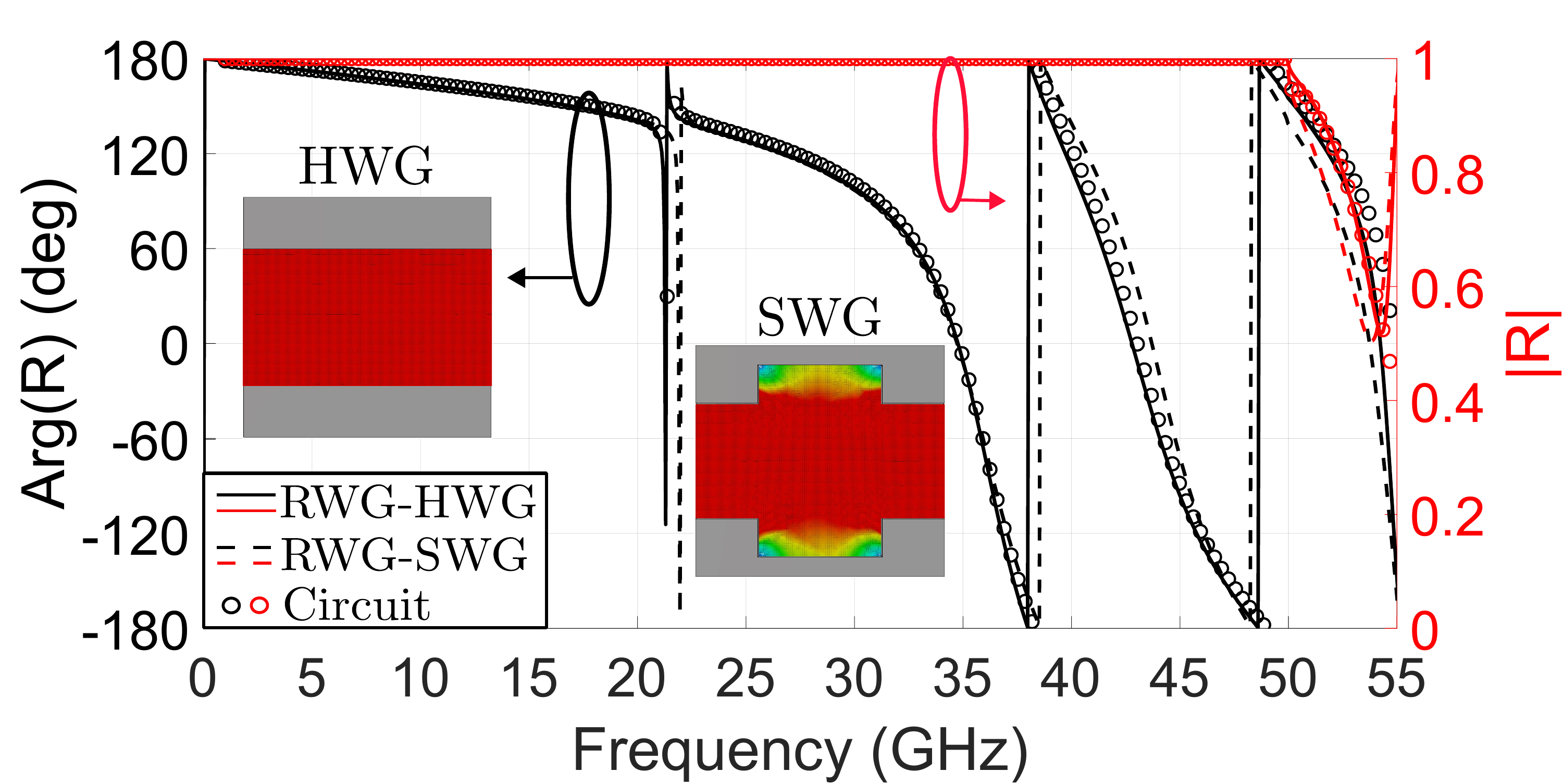}}
	\subfigure[]{\includegraphics[width= 1.05\columnwidth]{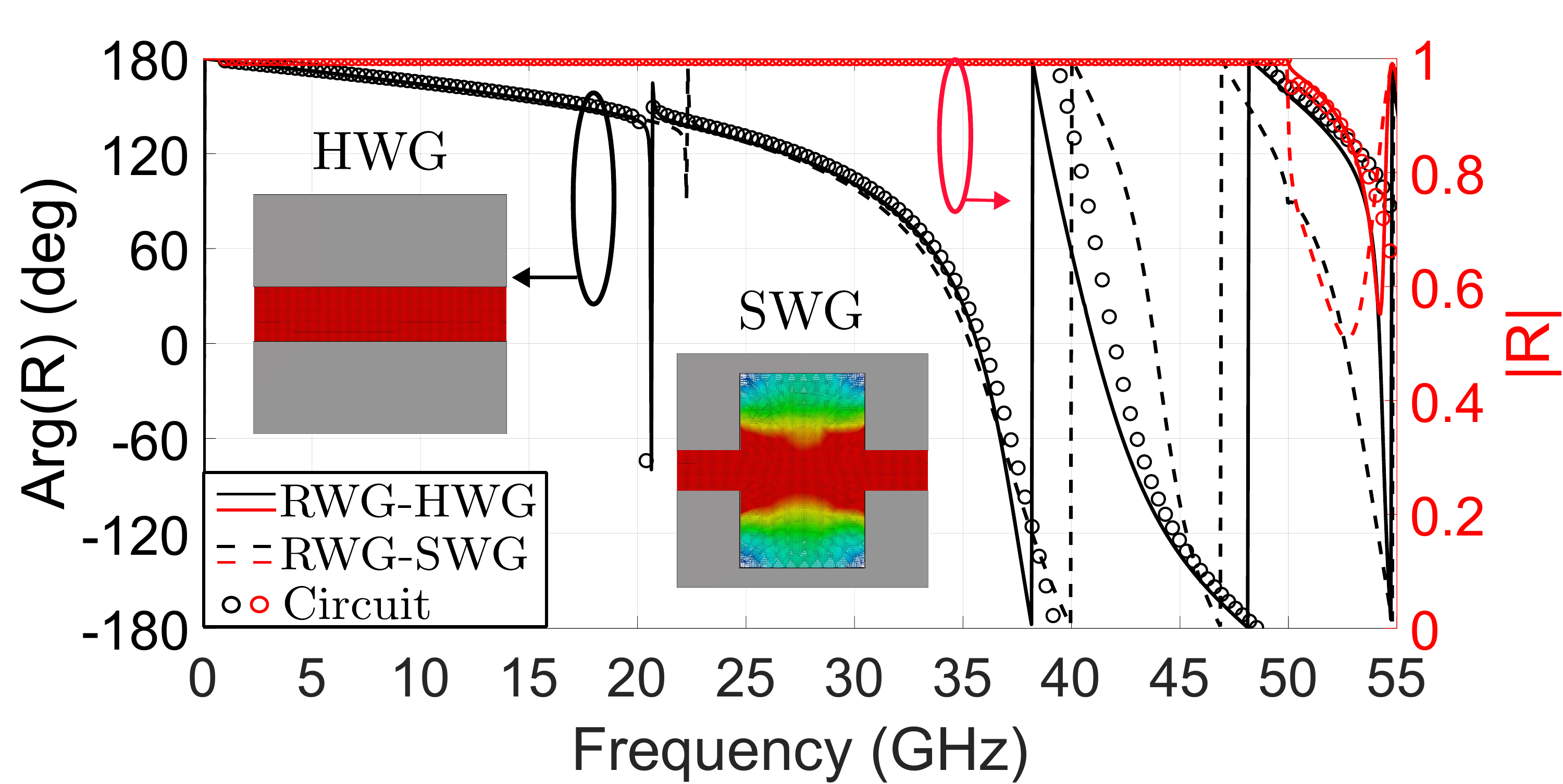}}
	\caption{Comparison of the phase of the reflection coefficient for two different reflective metastructures: original (RWG and HWG sections) and alternative (RWG and SWG sections). Colored solid and dashed lines correspond to full-wave results while colored circles correspond to analytical results. The height of the slotted regions is varied. (a) $h_{\text{SWG}}=3\,$mm, $h_{\text{HWG}}=3.42\,$mm. (b) $h_{\text{SWG}}=1\,$mm, $h_{\text{HWG}}=1.375\,$mm. The original and alternative 3D structures show similar electromagnetic responses when the slot height is large [panel (a)].  Geometrical parameters: $p_{\text{x}}=p_{\text{y}}=6\,$mm, $w_{\text{x}} = 5\,$mm, $w_{\text{y}} = 5\,$mm, $w_{\text{x}}^{\text{(SWG)}} = 3\,$mm, $d_1=4\,$mm, $d_2=6\,$mm.}
	\label{S11_comparacion}
\end{figure}

In general, electromagnetic fields within the SWG region are different and of more complex nature than of a HWG. Note that a HWG is bounded with perfect electric conditions at the upper and lower walls and magnetic conditions at the lateral walls, which simply excites a TEM mode at low frequencies. The situation is different in a SWG section, where the electric field profile shows a TEM-like profile within the slotted region but does not vanish outside this region. However, under certain circumstances, it can be demonstrated that the SWG can behave effectively as a HWG, i.e., $\mathbf{E}^{\text{(HWG)}} \approx \mathbf{E}^{\text{(SWG)}}$. In those cases, the original and alternative 3D metastructures will show similar electromagnetic responses and, therefore, both can be analyzed with the analytical equivalent circuit. This scenario is illustrated in \Fig{S11_comparacion}. The reflection coefficient $R$ is plotted for the original (solid lines) and alternative (dashed lines) 3D structures when varying the height of the longitudinal slots.   

\Figs{S11_comparacion}(a) and (b) present scenarios where wide and narrow longitudinal slots are considered, respectively. Results suggest that wide slots in the SWG \linebreak ($h_\mathrm{SWG} \gtrsim 0.6 w_{\text{y}}$) lead to similar electromagnetic responses between the original and alternative 3D metastructures. This is observed in a wide range of frequencies. On the other hand, the frequency range where the original and alternative 3D structures show a similar electromagnetic behavior significantly reduces when inserting narrow slots ($h_\mathrm{SWG} \lesssim 0.6 w_{\text{y}}$). This can be qualitatively explained by looking at the electric field profiles of the HWG and SWG shown in \Fig{S11_comparacion}. The electric field must vanish at the upper and lower metallic walls in the SWG region. In the extreme case $h_\mathrm{SWG} = w_{\text{y}}$, the SWG section transforms into an actual HWG governed by the fundamental TEM mode. For this extreme case, the electromagnetic response of both original and alternative 3D structures is indistinguishable.

As the slot gets progressively narrower ($h_\mathrm{SWG}$ decreases), differences between both configurations start to appear, mainly at high frequencies. In cases where the slot is wide [\Fig{S11_comparacion}(a)], the electric field in the SWG is mainly of TEM nature, as the edge of the slot is close to the upper and lower metallic walls. From a practical perspective, the complex SWG can be reduced to an equivalent HWG section in order to operate. The equivalent HWG would be of same width ($p_{\text{x}}$) than the SWG and effective height $h_\mathrm{eff}$. The concept of \emph{effective height} is heuristically introduced after inspecting the excited fields in the SWG. It can be observed that the electric field is fundamentally confined in the slot ($h_\mathrm{SWG}$) plus a tiny vertical region $\Delta h_\mathrm{SWG}$, thus $h_\mathrm{eff} = h_\mathrm{SWG} + \Delta h_\mathrm{SWG}$. By analyzing the SWG in the alternative 3D structure as an equivalent HWG, the circuit models is able to provide accurate results on the scattering phenomena, as \Fig{S11_comparacion}(a) illustrates.

The situation is not that simple when considering narrow SWG slot insertions ($h_\mathrm{SWG} \lesssim 0.6 w_{\text{y}}$). In this case, as \Fig{S11_comparacion}(b) illustrates, the excited fields in the SWG departs from the TEM profile seen in the HWG. This fact becomes especially noticeable at high frequencies, where the original (solid lines) and alternative (dashed lines) 3D structures no longer show similar electromagnetic responses. Consequently, as discussed at the end of subsection \ref{transmitting}, the   results obtained with the circuit (circles)  and CST (solid line) slightly differ from each other when considering narrow slots. This is due to the intense capacitive effects introduced by $Y_\mathrm{eq}^\mathrm{II}$ in the HWG, thus leading to a resonant frequency extracted by CST ($38.16$ GHz) to be somewhat lower than that calculated by the circuit ($39.26$ GHz).  Distance between the top and bottom walls at the center of the SWG region (set by $w_{\text{y}}$) provokes the capacitance of the SWG to be smaller than of the HWG. 

\begin{figure}[t!]
	\centering
    \includegraphics[width= 1.09\columnwidth]{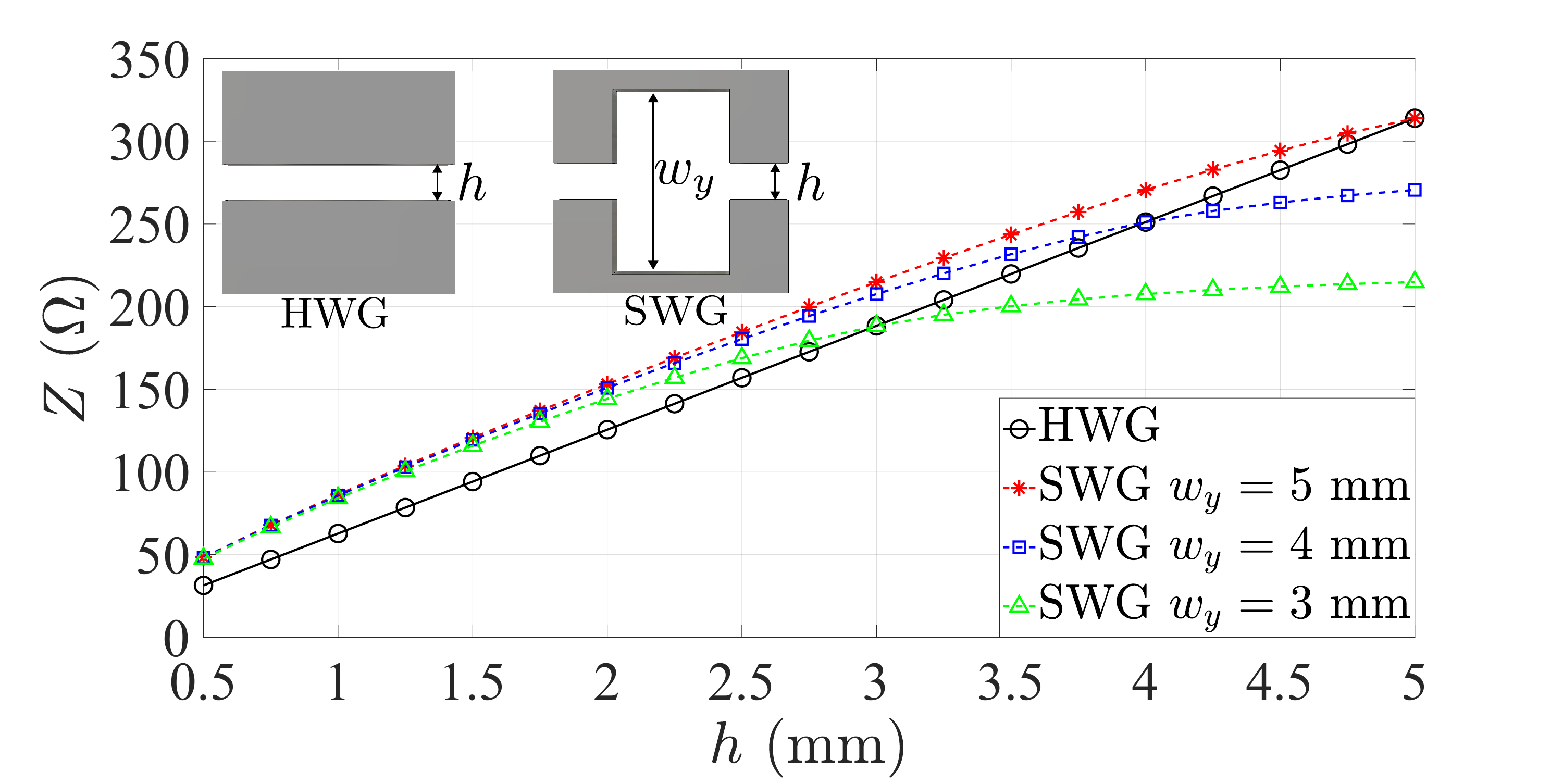}
	\caption{Line impedance in the the SWG and HWG as a function of the height $h$. Geometrical parameters: $p_{\text{x}}=p_{\text{y}}=6\,$mm, $w_{\text{x}}^{\text{(SWG)}} = 3\,$mm. Results extracted with CST Studio.}
	\label{Z_vs_H}
 \end{figure}

In order to estimate the values for the effective height $h_\mathrm{eff}$, \Fig{Z_vs_H} depicts the line impedance at $30$ GHz of both HWG and SWG structures. Note that the line impedance is plotted as a function of the height of the slot ($h$ parameter). These results have been extracted from port information in commercial software CST (v2022). It is worth noting that $Z$ is the impedance associated with the propagative TEM mode through the uniform line. In the SWG, these curves has been extracted for three different values of $w_{\text{y}}$: $w_{\text{y}} = 3$ mm (dashed green line), $w_{\text{y}} = 4$ mm (dashed blue line) and $w_{\text{y}} = 5$ mm (dashed red line).

Naturally, when the values of $w_{\text{y}}$ and $h$ are identical in the SWG structure, the extracted impedance coincides with that of the HWG. As a result, the impedance curves cross each other. However, the impedance values that are of particular interest in this work are those above the HWG curve (black solid line). In these cases, the original and alternative 3D structures are expected to show similar electromagnetic responses and, therefore, both can be analyzed with the present analytical circuit approach. Thus, it can be seen that for any selected height of SWG, there is another (greater) effective height of the HWG where their impedances coincide. As an example, a SWG region with $h=3$ mm and $w_{\text{y}}=5$ mm will present the same line impedance (at 30 GHz) than a HWG region with $h_{\mathrm{eff}}=3.42$ mm.

\section{Applications: Full-Metal Polarizer}

\begin{figure}[!h]
	\centering
    \subfigure[]{\includegraphics[width= 0.8\columnwidth]{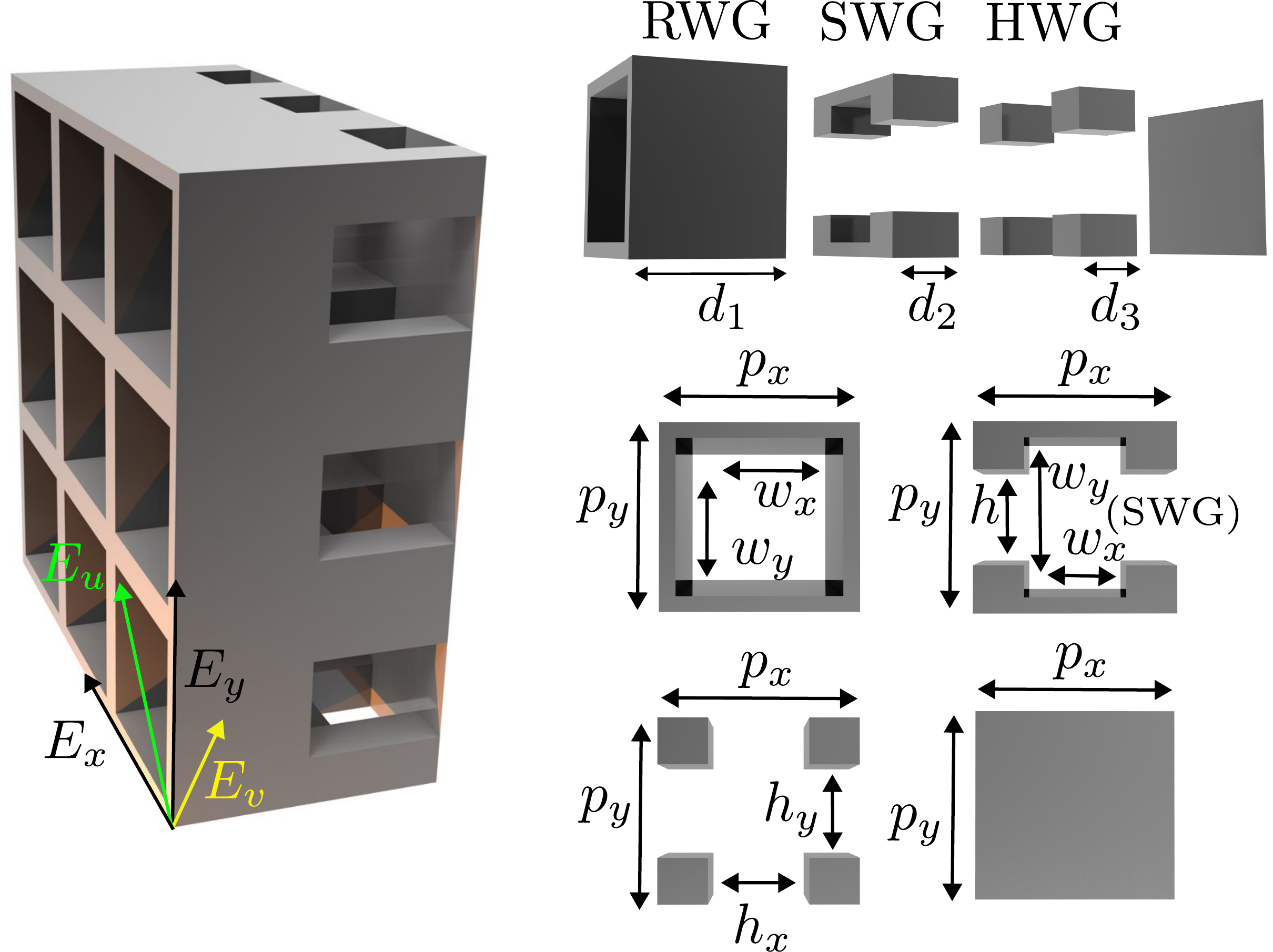}}
	\subfigure[]{\includegraphics[width= 1\columnwidth]{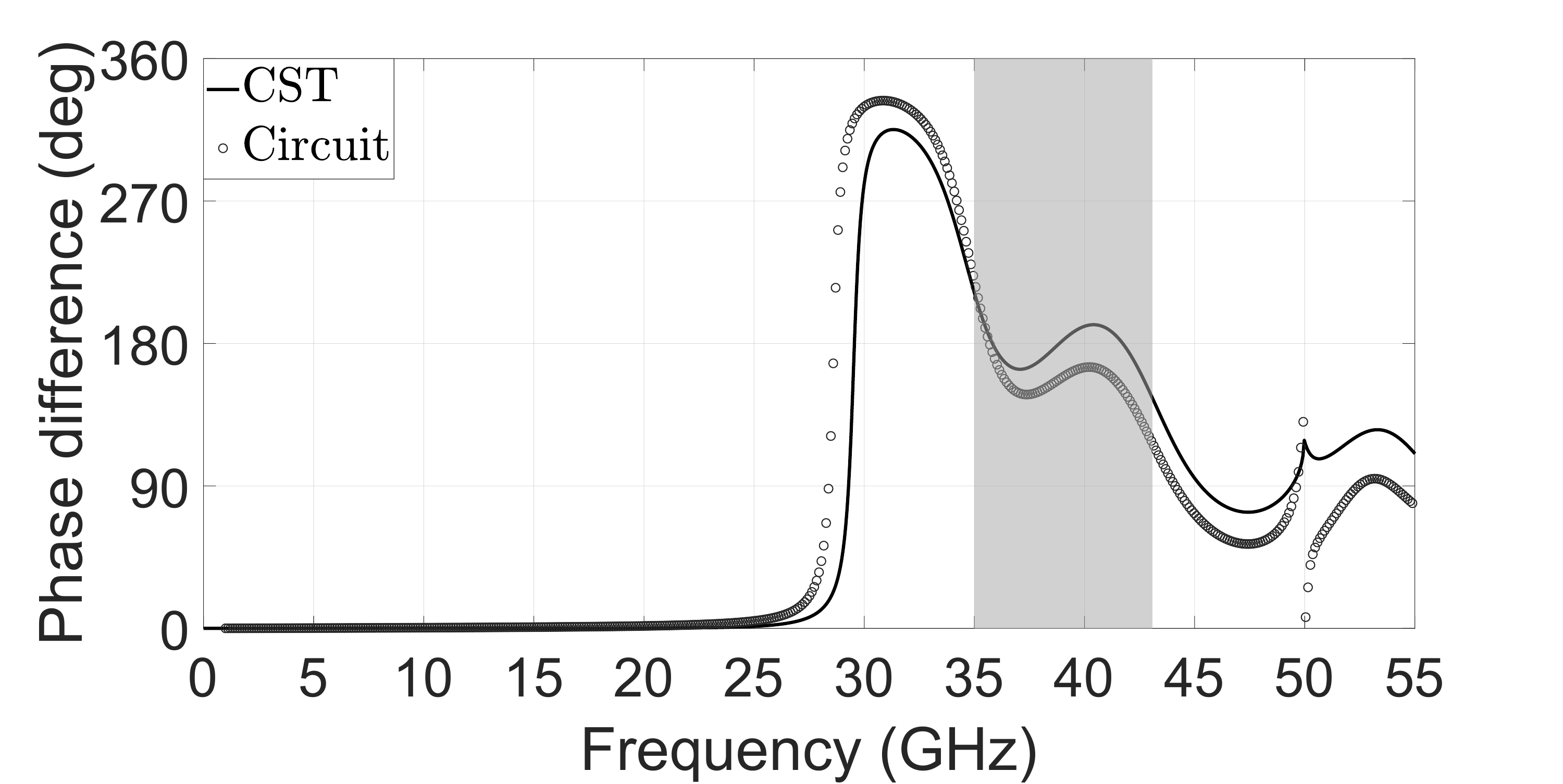}} 
    \subfigure[]{\includegraphics[width= 1\columnwidth]{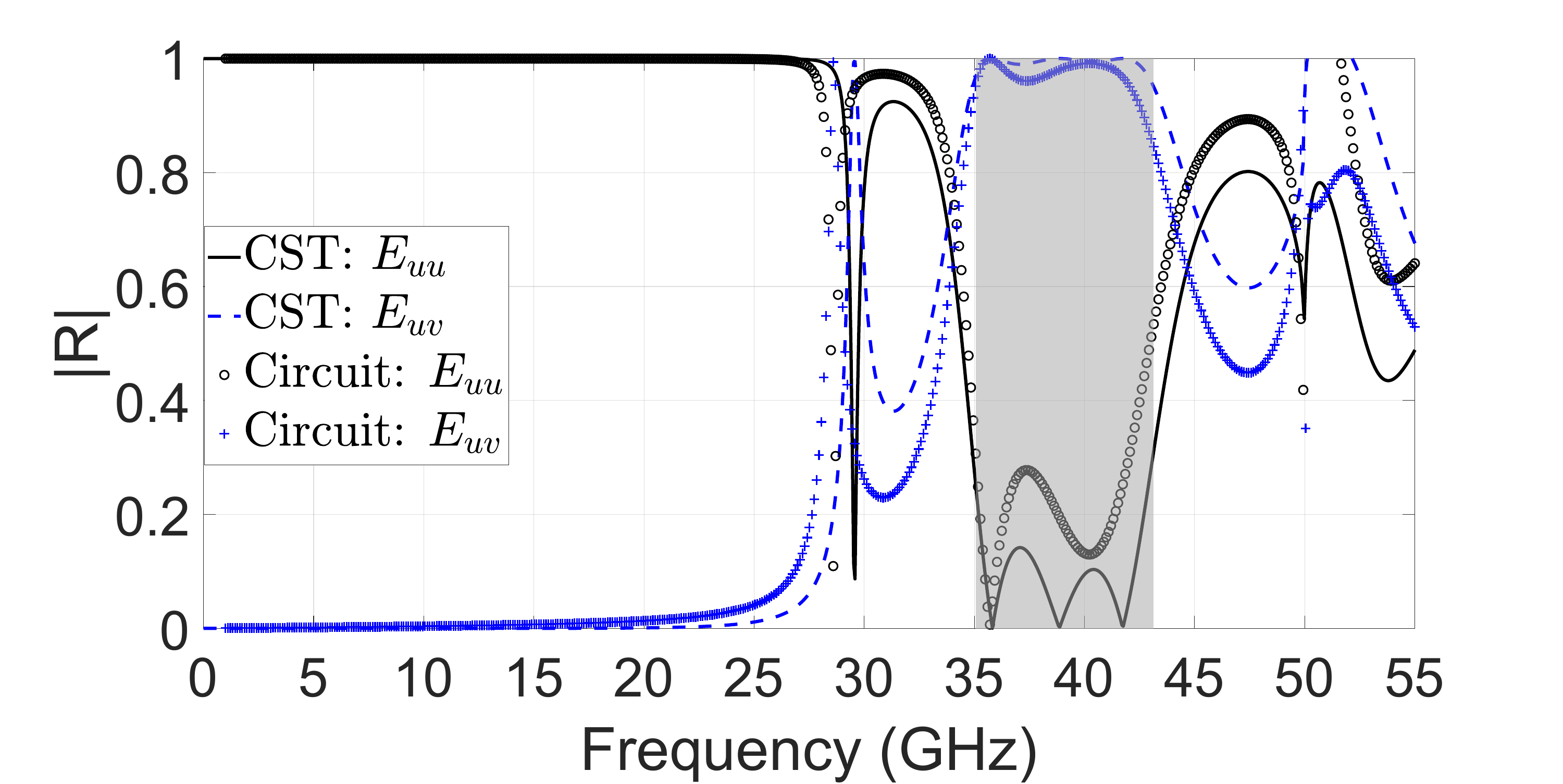}} 
	\caption{(a) Illustration of a 3D reflective polarization rotator.  (b) Phase difference between the reflection coefficient of the vertical co-polarization ($E_{yy}$) and the reflection coefficient of the horizontal  co-polarization ($E_{xx}$). (c) Amplitude of the reflection coefficient $R$, showing the co-pol. (black) and cross-pol. (blue) components for $E_u$ and $E_v$. The shadowed region (in grey) indicates the operative range of the polarization rotator. Solid and dashed lines correspond to full-wave results, while circles and crosses to analytical ones. Geometrical parameters: $p_{\text{x}}=p_{\text{y}}=6\,$mm, $w_{\text{x}} = w_{\text{y}}= 5\,$mm, $w_{\text{x}}^{\text{(SWG)}} = 3\,$mm, $d_1=4\,$mm, $d_2=d_3=2\,$mm, $h=3\,$mm, $h_{\text{x}}=h_{\text{y}}=3\,$mm.}
	\label{Polarizador}
\end{figure}

One of the main advantages of the proposed 3D metamaterials is the independent control of the two orthogonal polarization states. This fact has become evident after evaluating the results obtained in \Figs{PPWpartido} and \ref{S11_comparacion}. Independent polarization control is a remarkable feature, rarely found in 1D and 2D implementations, that can be advantageously exploited for the design and prototyping of full-metal polarizers. 

As an example, the proposed analytical approach is applied for the design of a reflective polarizer that operates from 35 GHz to 43.2 GHz. The proposed 3D device rotates the polarization plane of the reflected wave 90 degrees in the selected frequency range. Consequently, the direction of rotation of a circularly polarized incident wave would also change. This is one of the main applications in which the analytical circuit study can be used from an engineering perspective. It is worth noting that the purpose of this section is not to design a state-of-the-art polarizer, but to show the applicability of the circuit approach in real-world scenarios. 

\Fig{Polarizador}(a) sketches the fully-metallic 3D reflective polarizer.  As it can be appreciated, the 3D polarizer is composed of all the regions depicted above: RWG, SWG and HWG. Note that, with this configuration it is possible to achieve slots of different lengths in the vertical and horizontal walls, thus the resonances can be tuned by modifying parameters $d_2$ and $d_3$.

Reflective fully-metallic 3D structures allow simple but efficient polarizer designs due to the low ohmic losses involved. As the polarizers are short-circuited at the end of the structures, only the phase shift between the orthogonal components of the incident electric field needs to be set for the design of polarizers of different nature. In order to polarize an incoming wave whose electric field vibrates along $\hat{\mathbf{u}}$-direction [shown in \Fig{Polarizador}(a)], it is sufficient to set the phases separately from the reflection coefficient $R$ of its main components $E_{\text{x}}$ and $E_{\text{y}}$. Therefore, if the reflection coefficient of $E_{\text{x}}$ and $E_{\text{y}}$ in their co-polar direction ($E_{\text{xx}}$ and $E_{\text{yy}}$) present a phase shift of 180 deg, a rotator is achieved. From a design perspective, it is common to assume a maximum error of $180 \,\textrm{deg} \pm 37\, \textrm{deg}$ \cite{Balanis2015, Zhang2021}. This error band fixes the operation bandwidth of the polarization rotator. 


The operation of the polarizator rotator is as follows. In both $E_{\text{xx}}$ and $E_{\text{yy}}$ components, the incident wave impinges in the RWG section whose dimensions prevent the propagation of the fundamental mode below 30 GHz. Subsequently, the propagative $\text{TE}_{10}^{(\mathrm{RWG})}$ mode is guided from $30$ GHz to $50$ GHz by the SWG and the short-circuited HWG as a TEM-like and TEM mode respectively. Thus, single-mode operation is guaranteed from 30 GHz to 50 GHz. Above 50 GHz,  the amplitude decays as the second propagative mode is excited. A center frequency of 37.4 GHz is assumed as a starting design point. At this frequency, the shorter vertical slot insertion (SWG-HWG-short section) creates a $\lambda/4$ resonance, while the larger horizontal slot insertion (HWG-short section) creates a $\lambda/2$ resonance. Both $\lambda/4$ and $\lambda/2$ resonances produce a relative phase shift of 180 deg between the horizontal and vertical components, which leads to a rotation of the polarization plane for the reflected wave.
The study realized in section III has made it possible to simulate the SWG region as a HWG with parameters that approximate their electromagnetic responses.

\Fig{Polarizador}(b) illustrates the phase difference of the reflection coefficient $R$ of the two orthogonal components $E_{\text{xx}}$ and $E_{\text{yy}}$. The shadowed region indicates the simulated bandwidth where the 3D polarization rotator is operative. In \Fig{Polarizador}(c), the rotation on the electric field is directly visualized. It shows the amplitude of the reflection coefficient $R$ for an incident wave whose electric field vibrates according to $E_u$. The shadowed region sets the previous bandwidth where the co-polarization reflection is below to -10 dB in logarithmic scale. Consequently, the power is transmitted to the orthogonal component $E_v$. As can be seen, the bandwidth obtained by the circuit model is slightly lower than the simulated bandwidth. Despite this, a good agreement is observed between full-wave numerical results in CST and the analytical circuit taking into account the variations from the original structure shown in \Fig{fig1}. Naturally, it is important to remark the difference in computational times between CST and the circuit in the design stage, beyond the physical insight that the circuit may also provide. For the extraction of the two orthogonal components, CST took approximately $10$ minutes and the analytical circuit only 2 seconds.

It is interesting to highlight the reduced weight of the 3D polarizer. As the polarizer is based on the use of periodic waveguides (RWG, HWG, SWG), the interior of the structure is essentially hollow and filled with air. This causes the weight of the structure to be much less than originally suspected, since the volume of metal (or metallized material) is small compared to the total volume. Of the total volume of the polarizer, approximately 63\% is air while only 37\% is metal. This leads to a reduced weight for the polarizer, as well as for the others 3D metastructures presented in this work, which is of great interest for potential commercial applications. 

Additionally, the size of the proposed 3D polarizer is smaller than one may initially think. The dimensions of its unit cell are $0.75\lambda \times 0.75\lambda \times \lambda$ (width $\times$ height $\times$ thickness), considering a central frequency of 37.5 GHz. A fabricated functional polarizer would require, at least, of a structure formed by $10\times 10$ unit cells so the real-world finite implementation can be approached as periodic and analyzed with the present mathematical framework. Thus, the dimensions of the fabricated finite structure would be $7.5\lambda \times 7.5\lambda \times \lambda$. If we compare these dimensions with those required in a lens-type antenna or with other reflectarrays/transmitarrays based on 2D configurations, we see that the 3D polarizer has a relatively compact size. A compact size is a great practical advantage when fabricating and using the 3D metastructures, e.g. for integration into mobile communication platforms.



\section{\label{sec:Conc} Conclusion}

Analytical modeling of complex 3D metastructures has been elusive due to the intricate geometries involved. In this paper, we have derived an analytical framework for the analysis of 3D metamaterials formed by periodic arrangements of rectangular waveguides with longitudinal slot insertions. The proposed approach comes with an associated analytical equivalent circuit. The analytical circuit model is constituted by transmission lines that model wave propagation through the different homogeneous waveguide sections and shunt equivalent admittances that model higher-order harmonic excitation at the discontinuities. It is shown that the slotted waveguide sections can be modeled as a general waveguide with periodic boundary conditions (PBC). For the selected input excitation, and especially for normal incidence, PBC conditions can be relaxed to perfect magnetic conductor (PMC) conditions. Thus, slotted waveguides sections can be modeled as ``hard waveguides" (HWG).  We have tested the analytical framework against full-wave numerical results. A good agreement is observed in all cases, when considering reflective (short circuited) and transmitting (open) 3D configurations in normal and oblique incidence conditions. In the 3D metagrating, narrowband transmission is observed at frequencies below the cutoff. In cases where the rectangular waveguide (RWG) is short, evanescent waves may couple to the HWG (led by its fundamental TEM mode) and transmit without losses along the 3D structure. The analytical circuit model has proven to be a powerful tool to gain physical insight into complex scattering and diffraction phenomena, such as this one. Additionally,  results show that the proposed 3D metamaterial is suitable for the efficient design of full-metal polarizer devices with advanced or complex functionalities. This is attributed to the reliable independent polarization control of the two orthogonal states that the 3D metamaterial exhibits. An example is the polarization rotator shown in the work, which operates from 35 to 44 GHz approximately. The good agreement exhibited by full-wave and circuit results makes the circuit approach to become a rapid design tool. We hope that the analytical methodology developed in this work will be a step forward in the analysis and design of more advanced and, surely, more attractive 3D metastructures.

\begin{acknowledgments}
This work has been supported by grant PID2020-112545RB-C54 funded by MCIN/AEI/10.13039/501100011033 and by the European Union NextGenerationEU/PRTR. It has also been supported by grants PDC2022-133900-I00, TED2021-129938B-I00 and TED2021-131699B-I00.

\end{acknowledgments}


\section*{A. Transmission Matrices}
In the way that the analytical circuit model is defined in \Fig{fig1}(c), the computation of the scattering parameters is based on the usage of transmission (ABCD) matrices. Shunt  equivalent admittances $Y_\mathrm{eq}$ can be modeled with the following ABCD matrix \cite{pozar}
\begin{equation}
 \big[\mathrm{\mathbf{T}}_{Y}(Y_\mathrm{eq})\big] = 
\begin{bmatrix}
1 & 0\\
Y_\mathrm{eq} & 1
\end{bmatrix} \,.
\end{equation} 
Note that a short circuit in reflective structures can be modeled with the previous expression by simpling selecting $Y_\mathrm{eq} \rightarrow \infty$. Waveguide sections (RWG and HWG) are modeled as lossless transmission lines according to \cite{pozar}
\begin{equation}
\big[\mathrm{\mathbf{T}}_{L}(Y_0, \beta, d)\big] = 
\begin{bmatrix}
\cos(\beta d) & \jj \frac{1}{Y_0} \sin(\beta d)\\
\jj Y_0 \sin(\beta d) & \cos(\beta d)
\end{bmatrix},\quad
\end{equation}
where $Y_0$, $\beta$ and $d$ are the characteristic admittance, propagation constant and length of the transmission line. 

The global transmission matrix $\big[\mathrm{\mathbf{T}}\big]$ is computed by cascading (multiplying) the individual transmission matrices that form the complete circuit. Then, the global scattering matrix $\big[\mathrm{\mathbf{S}}\big]$ is extracted from $\big[\mathrm{\mathbf{T}}\big]$ by using the formulas detailed in \cite{Frickey1994}, which take into consideration that the input and output media (impedance/admittance of the reference ports) could be different.

\bibliography{./ref}

\end{document}